\documentclass[pre,twocolumn,amsmath,amssymb]{revtex4}

\newtheorem{definition}{Definition}

\newtheorem{thm}{Theorem}
\usepackage{graphicx}
\usepackage{amsmath}

\begin{document}

\title{A Guided Walk Down Wall Street: an Introduction to Econophysics}

\author{Giovani L.~Vasconcelos}
\email{giovani@lftc.ufpe.br}
\affiliation{Laborat\'orio de F\'\i sica Te\'orica e Computacional, 
Departamento de F\'{\i}sica, Universidade Federal de Pernambuco,
50670-901 Recife, PE, Brazil\\ \\
}
\date{May 10, 2004\\}

\begin{abstract}
\vspace{0.5cm}
This article contains the lecture notes for the short course
``Introduction to Econophysics,'' delivered at the II Brazilian School
on Statistical Mechanics, held in S\~ao Carlos, Brazil, in February
2004.  The main goal of the present notes is twofold: i) to provide a
brief introduction to the problem of pricing financial derivatives in
continuous time; and ii) to review some of the related problems to
which physicists have made relevant contributions in recent years.
\end{abstract}


\maketitle

\section{Introduction}

This article comprises the set of notes for the short course
``Introduction to Econophysics,'' delivered at the II Brazilian School
on Statistical Mechanics, held at the University of S\~ao Paulo, in
S\~ao Carlos, SP, Brazil, in February 2004. The course consisted of
five lectures and was aimed at physics graduate students with no
previous exposition to the subject.

The main goal of the course was twofold: i) to provide a brief
introduction to the basic models for pricing financial derivatives;
and ii) to review some of the related problems in Finance to which
physicists have made significant contributions over the last decade.
The recent body of work done by physicists and others have produced
convincing evidences that the {\it standard model of Finance} (see
below) is not fully capable of describing real markets, and hence new
ideas and models are called for, some of which have come straight from
Physics. In selecting some of the more recent work done by physicists
to discuss here, I have tried to restrict myself to
problems that may have a direct bear on models for pricing
derivatives.  And even in such cases only a brief overview of the
problems is given. It should then be emphasized that these notes are
not intended as a review article on Econophysics, which is nowadays a
broad interdisciplinary area, but rather as a pedagogical introduction
to the mathematics (and physics?)  of financial derivatives.  Hence no
attempt has been made to provide a comprehensive list of references.

No claim of originality is made here regarding the contents of the
present notes. Indeed, the basic theory of financial derivatives can
now be found in numerous textbooks, written at a different
mathematical levels and aiming at specific (or mixed) audiences, such as
economists \cite{hull,duffie,ingersoll,bjork}, applied
mathematicians \cite{shiryaev,wilmott,mikosch,oksendal}, physicists
\cite{MS,BP,voit}, etc. (Here I have listed only the texts that were
most often consulted while writing these notes.)  Nevertheless, some
aspects of presentation given here have not, to my knowledge, appeared
before.  An example is the analogy between market efficiency and a
certain symmetry principle that is put forward in Sec.~V.  Similarly,
the discussion of some of the more recent research problems is based
on the already published literature.  An exception is
Fig.~\ref{fig:multi} which contains unpublished results obtained by
R.~L.~Costa and myself.

The present notes are organized as follows. Section II gives some
basic notions of Finance, intended to introduce the terminology as
well as the main problems that I shall be considering. In Sec.~III, I
discuss the Brownian motion, under a more formal viewpoint than most
Physics graduate students are perhaps familiar with, and then develop
the so-called It\^o stochastic calculus.  Section IV contains what is
the {\it raison d'etre} of the present notes, the Black-Scholes model
for pricing financial derivatives. In Sec.~V, the martingale approach
for pricing derivatives is introduced.  In particular, I recast the
notions of market efficiency and no-arbitrage as a `symmetry
principle' and its associated `conservation law.'  Sections VI and VII
discuss two possible ways in which real markets may deviate from the
standard Black-Scholes model.  The first of such possibilities is that
financial asset prices have non-Gaussian distributions (Sec.~VI),
while the second one concerns the presence of long-range correlations
or memory effects in financial data (Sec.~VII). Conclusions are
presented in Sec.~VIII. For completeness, I give in Appendix A the
formal definitions of probability space, random variables, and
stochastic processes.

\section{Basic Notions of Finance}

\subsection{Riskless and risky financial assets}

Suppose you deposit at time $t=0$ an amount of R\$ 1 into a bank
account that pays an interest rate $r$. Then over time the amount
of money you have in the bank, let us call it $B(t)$, will increase at
a rate
\begin{equation}
\frac{dB}{dt}=rB. \label{eq:dB}
\end{equation}
Solving this equation subject to the initial condition $B(0)=1$ yields
\begin{equation}
B(t)=e^{rt}. \label{eq:Bt}
\end{equation}

A bank account is an example of a riskless financial assets, since you
are guaranteed to receive a known (usually fixed) interest rate $r$,
regardless of the {\it market situation}. Roughly speaking, the way
banks operate is that they borrow from people who have money to
`spare', but are not willing to take risks, and lend (at higher
interest rates) to people who `need' money, say, to invest in some
risky enterprise.  By diversifying their lending, banks can reduce
their overall risk, so that even if some of these loans turn bad they
can still meet their obligations to the investors from whom they
borrowed.

Governments and private companies can also borrow money from investors
by issuing {\it bonds}. Like a bank account, a bond pays a (fixed or
floating) interest rate on a regular basis, the main difference being
that the repayment of the loan occurs only at a specified time, called
the bond maturity.  Another difference is that bonds are not strictly
risk-free assets because there is always a chance that the bond issuer
may default on interest payments or (worse) on the principal.
However, since governments have a much lower risk to default than
corporations, certain government bonds can be considered to be risk
free.

A company can also raise capital by issuing {\it stocks} or {\it
shares}. Basically, a stock represents the ownership of a small piece
of the company.  By selling many such `small pieces', a company can
raise capital at lower costs than if it were to borrow from a bank.
As will be discussed shortly, stocks are {\it risky} financial assets
because their prices are subjected to unpredictable fluctuations.  In
fact, this is what makes stocks attractive to {\it aggressive}
investors who seek to profit from the price fluctuations
by pursuing the old advice to ``buy low and sell high.''

The buying and selling of stocks are usually done in organized
exchanges, such as, the New York Stock Exchange (NYSE) and the S\~ao
Paulo Stock Exchange (BOVESPA).  Most stock exchanges have {\it
indexes} that represent some sort of average behavior of the
corresponding market. Each index has its own methodology. For example,
the {\it Dow Jones Industrial Average} of the NYSE, which is arguably
the most famous stock index, corresponds to an average over 30
industrial companies. The Ibovespa index of the S\~ao Paulo Stock
Exchange, in contrast, represents the present value of a hypothetical
portfolio made up of the stocks that altogether correspond to 80\% of
the trading volume.  Another well known stock index is the Standard \&
Poor's 500 (S\&P500) Index calculated on the basis of data about 500
companies listed on the NYSE.  [Many other risky financial assets,
such as, currency exchange rates, interest rates, and commodities
(precious metals, oil, grains, etc), are traded on organized markets
but these will not be discussed any further in the present notes.]

\subsection{The random nature of stock prices}

\begin{figure}
\begin{center}
\includegraphics*[width=.9\columnwidth]{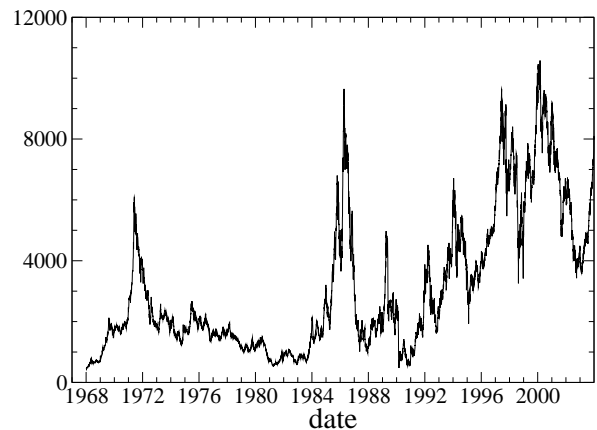}
\caption{Daily closing values of the deflated Ibovespa index 
in the period 1968--2003.} 
\label{fig:ibov}
\end{center}
\end{figure}

Since a stock represents a `small piece' of a company, the stock
price should somehow reflect the overall value (net worth) of this
company.  However,  the present value of a firm depends not only
on the firm's current situation but also on its {\it future
performance}.  So here one sees already the basic problem in pricing
risky financial assets: we are trying to predict the future on the
basis of present information. Thus, if a new information is revealed
that might in one way or another affect the company's future
performance, then the stock price will vary accordingly. It should
therefore be clear from this simple discussion that the future price
of a stock will always be subjected to a certain degree of
uncertainty. This is reflected in the typical `erratic behavior' that
stock prices show when graphed as a function of time. An example of
such a graph is shown in Fig.~\ref{fig:ibov} for the case of the
Ibovespa stock index.

Although stock prices may vary in a rather unpredictable way, this does
not mean that they cannot be modeled. It says only that they should
be described in a {\it probabilistic} fashion. To make the argument a little
more precise, let $S$ be the price of a given stock and suppose we
want to write an equation analogous to (\ref{eq:dB}) for the rate of
increase of $S$:
\begin{equation}
\frac{dS}{dt}=R(t)S, \label{eq:dS1}
\end{equation}
where $R(t)$ represents the `rate of return' of the stock. The
question then is: what is $R(t)$? From our previous discussion, it is
reasonable to expect that $R(t)$ could be separated into two
components: i) a predictable mean rate of return, to be denoted by
$\mu$, and ii) a fluctuating (`noisy') term $\xi(t)$, responsible for
the randomness or uncertainty in the stock price.  Thus, after writing
$ R(t)=\mu + \xi(t)$ in (\ref{eq:dS1}) we have
\begin{equation}
\frac{dS}{dt}=\left[\mu +\xi(t)\right] S. \label{eq:dS2}
\end{equation}
Now, one of the best models for `noise' is, of course, the white
noise, so it should not come as a surprise to a physicist that
Brownian motion and white noise play an important r\^ole in finance,
as will be discussed in detail shortly.

\subsection{Options and derivatives}

Besides the primary financial assets already mentioned (stocks,
commodities, exchange rate, etc), many other financial instruments,
such as {\it options} and {\it futures contracts}, are traded on
organized markets (exchanges). These securities are generically called
{\it derivatives}, because they derive their value from the price of
some primary underlying asset. Derivatives are also sometimes referred
to as {\it contingent claims}, since their values are contingent on
the evolution of the underlying asset. In the present notes, I will
discuss only one of the most basic derivatives, namely, options.

An option is a contract that gives its holder the right, {\it but not
the obligation}, to buy or sell a certain asset for a specified price
at some future time. The other part of the contract, the option
underwriter, is obliged to sell or buy the asset at the specified
price. The right to buy (sell) is called a {\it call} ({\it put})
option. If the option can only be exercised at the future date
specified in the contract, then it is said to be a European
option. American options, on the other hand, can be exercised at any
time up to maturity. (For pedagogical reasons, only European
derivatives will be considered here.) To establish some notation let
us give a formal definition of a European option.

\begin{definition}
A European call  option with exercise price (or strike price)
$K$ and maturity (or expiration date) $T$ on the underlying asset $S$
is a contract that gives the holder the right to buy  the
underlying asset for the price $K$ at time $T$.
\end{definition}

A European {\it put option} is the same as above, the only
difference being that it gives the holder the right to {\it sell} the
underlying asset for the exercise price at the expiration date.

\begin{figure}
\begin{center}
\includegraphics*[width=.9\columnwidth]{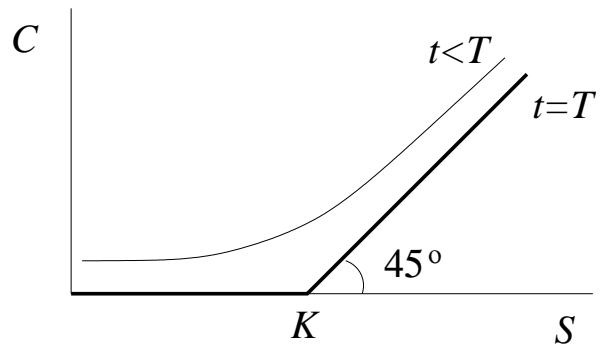}
\caption{Value of a call option at the expiration date (thick line) and before
expiration (thin line).} 
\label{fig:call1}
\end{center}
\end{figure}

\begin{figure}
\begin{center}
\includegraphics*[width=.9\columnwidth]{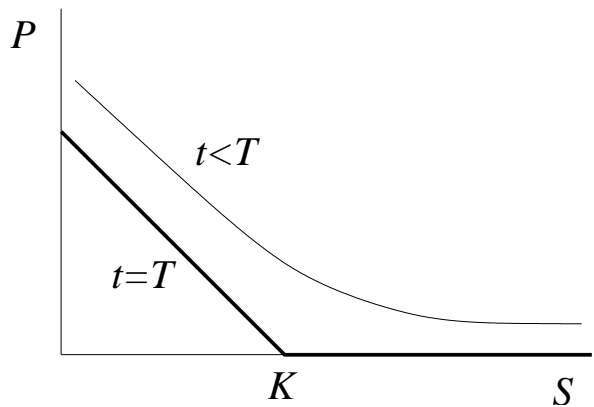}
\caption{Value of a put option at the expiration date (thick line) and before
expiration (thin line).} 
\label{fig:put1}
\end{center}
\end{figure}

If at the expiration date $T$ the stock price $S_T$ is above the strike price
$K$, the holder of a call option will exercise his right to buy the
stock from the underwriter at price $K$ and sell it in the market at
the spot price $S_T$, pocketing the difference $S_T-K$. On the other
hand, if at expiration the price $S_T$ closes below $K$ then the call
option becomes worthless (since it would be cheaper to buy the stock
in the market). The {\it payoff} of a call option at maturity is
therefore given by
\begin{equation}
{\rm payoff}_{\rm call} =\max(S_T-K,0). \label{eq:payoff}
\end{equation}
The payoff diagram of a call option is illustrated by the thick line
in Fig.~\ref{fig:call1}. In this figure the thin line represents the
price of the call option at an arbitrary time $t<T$ before expiration.
(The otpion price before expiration is always greater than the payoff
at expiration on account of the higher risks: the further way the
expiration date, the greater the uncertainty regarding the stock price
at expiration.)  Similarly, the payoff function for a put option is
\begin{equation}
{\rm payoff}_{\rm put} =\max(K-S_T,0), \label{eq:payoffP}
\end{equation}
which is shown as the thick line in Fig.~\ref{fig:put1}.

Because an option entitles the holder to a certain right it commands a
premium. Said differently, since the underwriter has an obligation
(while the holder has only rights) he will demand a payment, to be
denoted by $C_0$, from the holder in order to enter into such a
contract. Thus, in the case of a call option, if the option is
exercised the holder (underwriter) will make a profit (loss) given by
$\max(S-K,0)-C_0$; otherwise, the holder (underwriter) will have lost
(won) the amount $C_0$ paid (received) for the option.  And similarly
for a put option. Note then that the holder and the underwriter of an
option have opposite views regarding the direction of the market. For
instance, the holder of a call option is betting that the stock price
will increase (past the exercise price), whereas the underwriter hopes
for the opposite.

Now, given that the holder and the underwriter have opposite views as
to the direction of the market, how can they possibly agree on the
price for the option? For if the holder (underwriter) suspects that
the option is overvalued (undervalued) he will walk away from the
contract.  The central problem in option pricing is therefore to
determine the {\it rational price} price $C_0$ that ensures that
neither part `stands a better chance to win.'

A solution to this problem (under certain assumptions) was given in
1973 in the now-famous papers by Black and Scholes \cite{BS} and
Merton \cite{merton}, which won Scholes and Merton the Nobel prize in
Economics in 1997. (Black had died meanwhile.)  The history of options
is however much longer. In fact, the first scientific study of options
dates back to the work by the French mathematician Bachelier in 1900
\cite{bachelier}, who solved  the option pricing
problem above but under slightly wrong assumptions; see, e.g.,
\cite{voit} for a detailed discussion of Bachelier's work.

After an option (traded on exchange) is first underwritten, it can
subsequently be traded and hence its `market price' will be determined
by the usual bid-ask auction. It is nonetheless important to realize
that investors in such highly specialized market need some basic
pricing theory to rely on, otherwise investing in options would be a
rather wild (and dangerous) game. Indeed, only after the appearance of
the Black-Scholes model [and the establishment of the first option
exchange in Chicago also in 1973] have option markets thrived.  One of
the main objectives of the present notes is to explain the theoretical
framework, namely, the Black-Scholes model and some of its extensions,
in which options and other derivatives are priced. I will therefore
not say much about the practical aspects of trading with options.

\subsection{Hedging,  speculation, and arbitrage}
\label{sec:HSA}

Investors in derivative markets can be classified into three main
categories:  {\it hedgers}, {\it speculators}, and {\it arbitrageurs}.

{\it Hedgers} are interested in using derivatives to reduce the risk
they already face in their portfolio. For example, suppose you own a
stock and are afraid that its price might go down within the next
months. One possible way to limit your risk is to sell the stock now
and put the money in a bank account. But then you won't profit if the
market goes up. A better {\it hedging} strategy would clearly be to
buy a put option on the stock, so that you only have to sell the stock
if it goes below a certain price, while getting to keep it if the price goes
up. In this case an option works pretty much as an insurance: you pay
a small price (the option premium $C_0$) to insure your holdings
against possibly high losses.

{\it Speculators}, in contrast to hedgers, seek to make profit by
taking risks. They `take a position' in the market, by betting that
the price on a given financial asset will go either up or down. For
instance, if you think that a certain stock will go up in the near
future, you could ``buy and hold'' the stock in the hope of selling it
later at a profit. But then there is the risk that the price goes
down. A better strategy would thus be to buy a call option on the
stock. This not only is far cheaper than buying the stock itself but
also can yield a much higher return on your initial investment. (Why?)
However, if the market does not move in the way you expected and the
option expire worthless, you end up with a 100\% loss.  (That's why
speculating with option is a very risky business.)

{\it Arbitrageurs} seek to make a {\it riskless} profit by
entering simultaneously into transactions in two or more markets,
usually without having to make any initial commitment of money. The
possibility of making a riskless profit, starting with no money at
all, is called an arbitrage opportunity or, simply, an arbitrage.  A
more formal definition of arbitrage will be given later. For the time
being, it suffices to give an example of how an arbitrage opportunity
may arise.

But before going into this example, it is necessary first to discuss
the notion of a {\it short sell}.  `Shorting' means selling an
asset that one does not own.  For example, if you place an order to
your broker to short a stock, the broker will ``borrow'' a stock from
somebody else's account, sell it in the market, and credit the
proceeds into {\it your} account. When you then decide to close your
short position (there usually is a limit on how long an asset can be
held short), your broker will buy the stock in the market (taking the
money from your account) and return it to its original owner. If in
the meantime the stock prices decreased, the short sell brings a
profit, otherwise the short seller incurs in a loss. This is why a
short sell is usually done simultaneously with another operation to
compensate for this risk (as in the arbitrage example below).  It
should also be noted, in passing, that buying the actual asset
corresponds to taking a `long position' on this asset.

Let us now consider our hypothetical arbitrage example. Many
Brazilian companies listed in the S\~ao Paulo Stock Exchange also have
their stocks traded on the New York Stock Exchange in the form of the
so-called American Depository Receipt (ADR).  Suppose then that a stock
is quoted in S\~ao Paulo at R\$ 100, with its ADR counterpart trading
in New York at US\$ 34, while the currency rate exchange is 1 USD =
2.90 BRL.  Starting with no initial commitment, an arbitrageur could
sell short $N$ stocks in S\~ao Paulo and use the proceeds to buy $N$
ADR's in New York (and later have them transferred to S\~ao Paulo to
close his short position). The riskless profit in such operation would
be R\$ ($100-2.90\times 34)N= {\rm R}\$~1.40 \, N$.  (In practice, the
transaction costs would eliminate the profit for all but large
institutional investors \cite{hull}.)

Note, however, that such `mispricing' cannot last long: buy orders in
New York will force the ADR price up, while sell orders in S\~ao Paulo
will have the opposite effect on the stock price, so that an {\it
equilibrium price} for both the ADR and the stock is soon reached,
whereupon arbitrage will no longer be possible. In this sense, the
actions of an {\it arbitrageur} are self-defeating, for they tend to
destroy the very arbitrage opportunity he is acting upon---but before
this happens a lot of money can be made. Since there are many people
looking for such riskless chances to make money, a well-functioning
market should be  free of arbitrage. This is the main
idea behind the principle that in an {\it efficient market} there is
no arbitrage, which is commonly known as the ``no-free-lunch''
condition.

\subsection{The no-arbitrage principle in a (binomial) nutshell}
\label{sec:bin}

\begin{figure}[t]
\begin{center}
\includegraphics*[width=.8\columnwidth]{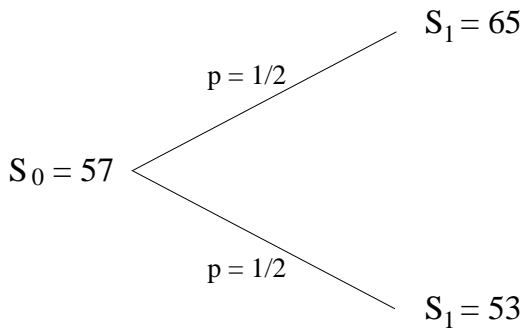}
\caption{One-step binomial model for a stock.} 
\label{fig:binS}
\end{center}
\end{figure}

\begin{figure}[t]
\begin{center}
\includegraphics*[width=.9\columnwidth]{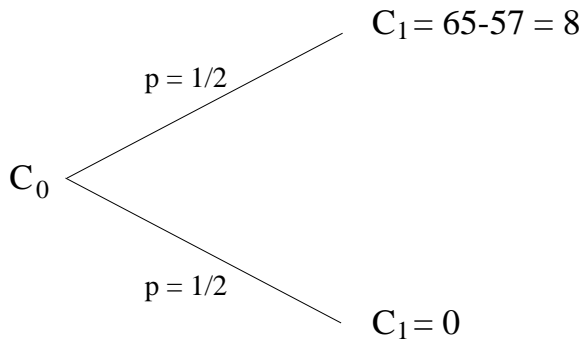}
\caption{Option value in the one-step binomial model.} 
\label{fig:binC}
\end{center}
\end{figure}

\begin{figure}
\begin{center}
\includegraphics*[width=.9\columnwidth]{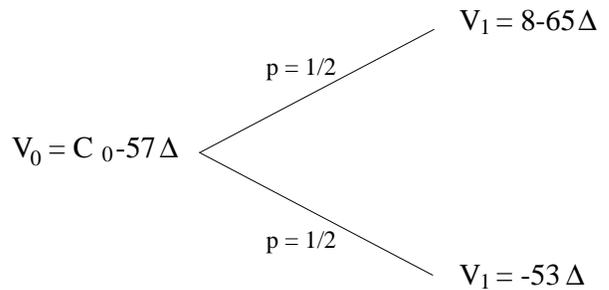}
\caption{Delta-hedging portfolio in the one-step binomial model.} 
\label{fig:binV}
\end{center}
\end{figure}

Here we shall consider a one-step binomial model to illustrate the
principle of no-arbitrage and how it can be used to price
derivatives. Suppose that today's price of an ordinary Petrobras
stocks (PETR3 in their Bovespa acronym) is $S_0 = 57$ BRL. Imagine
then that in the next time-period, say, one month, the stock can
either go up to $S_1^u=65$ with probability $p$ or go down to
$S_1^d=53$ with probability $q$. For simplicity let us take $p=q=1/2$.
Our binomial model for the stock price dynamics is illustrated in
Fig.~\ref{fig:binS}.  Note that in this case the stock mean rate of
return, $\mu$, is given by the expression: $(1+\mu)S_0=E[S_1]$, where
$E[S]$ denotes the expected value of $S$ (see Sec.~\ref{sec:1DBM}
for more on this notation). Using the values shown in
Fig.~\ref{fig:binS}, one then gets $\mu= 0.035$ or $\mu=3.5\%$. Let us also
assume that the risk-free interest rate is $r=0.6\%$ monthly.

Consider next a call option on PETR3 with exercise price $K=57$ and
expiration in the next time period, i.e., $T=1$. Referring to
(\ref{eq:payoff}) and Fig.~\ref{fig:binS}, one immediately sees that
at expiration the possible values (payoffs) for this option in our
binomial model are as shown in Fig.~\ref{fig:binC}: $C_1^u=8$ or $C_1^d=0$
with equal probability.  The question then is to determine the
`rational' price $C_0$ that one should pay for the option. Below we
will solve this problem using two different but related methods.  The
idea here is to illustrate the main principles involved in option
pricing, which will be generalized later for the case of continuous
time.

First, we describe the so-called delta-hedging argument.  Consider a
portfolio made up of one option $C$ and a short position on $\Delta$
stocks, where $\Delta$ is to be determined later, and let $V_t$ denote
the money value of such a portfolio at time $t$. We thus have
$$V_t=C_t-\Delta S_t,$$ where the minus sign denotes that we have
short sold $\Delta$ stocks (i.e., we `owe' $\Delta$ stocks in the
market).  From Figs.~\ref{fig:binS} and \ref{fig:binC}, one clearly
sees that the possibles values for this portfolio in our one-step
model are as illustrated in Fig.~\ref{fig:binV}.  Let us now chose
$\Delta$ such that the value $V_1$ of the portfolio is the same in
both `market situations.'  Referring to Fig.~\ref{fig:binV} one
immediately finds
\[ 
V_1^u=V_1^d\Longrightarrow 8-\Delta \cdot 65=-\Delta\cdot53 \Longrightarrow \Delta =\frac{2}{3}.
\] 
Thus, by choosing $\Delta=2/3$ we have {\it completely} eliminated the
risk from our portfolio, since in both (up or down) scenarios the
portfolio has the same value $V_1$.  But since this portfolio is
riskless, {\it its rate of return must be equal to the risk-free
interest rate $r$}, otherwise there would be an arbitrage opportunity,
as the following argument shows.

Let $r'$ denote the portfolio rate of return, i.e., $r'$ is the
solution to the following equation
\begin{equation}
(1+r')V_0=V_1. \label{eq:r'}
\end{equation}
If $r'<r$, then an {\it arbitrageur} should take a long position on
(i.e., buy) the option and a short position on $\Delta$ stocks. To see
why this is an arbitrage, let us go through the argument in detail.
At time $t=0$ the arbitrageur's net cashflow would be
$B_0=|V_0|=\Delta\cdot S_0-C_0$, which he should put in the bank so
that in the next period he would have $B_1=(1+r)|V_0|$. At time $t=1$,
he should then close his short position on $\Delta$ stocks, either
exercising his option (up scenario) or buying $\Delta$ stocks directly
in the market (down scenario). In either case, he would have to pay
the same amount $|V_1|=(1+r')|V_0|<B_1$, and hence would be left with
a profit of $B_1-|V_1|$. On the other hand, if $r'>r$ the arbitrageur
should adopt the opposite strategy: go short on (i.e., underwrite) the
option and long on $\Delta$ stocks (borrowing money from the bank to
do so).

We have thus shown that to avoid arbitrage we must have $r'=r$.  This
is indeed a very general principle that deserves to be stated in full:
{\it in a market free of arbitrage any riskless portfolio must yield
the risk-free interest rate $r$}. This no-arbitrage principle is at
the core of the modern theory of pricing derivatives, and, as such, it
will be used several times in these notes.

Let us now return to our option pricing problem.  Setting $r'=r$ in
(\ref{eq:r'}) and substituting the values of $V_0$ and $V_1$ given in
Fig.~\ref{fig:binV}, we obtain
\begin{equation}
(1+r)\left[C_0-\Delta \, S_0\right]=-\Delta \, S_1^d.
\end{equation}
Inserting the values of $r=0.006$, $S_0=57$, $S_1^d=53$, and
$\Delta=2/3$ into the equation above, it then follows that the option
price that rules out arbitrage is
\begin{equation}
C_0=2.88. \label{eq:C0}
\end{equation}

It is instructive to derive the option price through a second method,
namely, the {\it martingale approach} or {\it risk-neutral
valuation}. To this end, we first note that from Fig.~\ref{fig:binC}
we see that the expected value of the option at expiration is
$E[C_1]=\frac{1}{2} \; 8 + \frac{1}{2} \; 0 = 4$. One could then
think, not totally unreasonably, that the correct option price should
be the expected payoff discounted to the present time with the
risk-free interest rate. In this case one would get
$$C_0'=\displaystyle\frac{E[C_1]}{1+r}=\frac{4}{1.006}=3.98,$$ which
is quite different from the price found in (\ref{eq:C0}). The faulty
point of the argument above is that, while we used the risk-free rate
$r$ to discount the expected payoff $E[C_1]$, we have implicitly used
the stock mean rate of return $\mu$ when calculating $E[C_1]$.  Using
these two different rates leads to a wrong price, which would in turn
give rise to an arbitrage opportunity.

A way to avoid this arbitrage is to find fictitious probabilities
$q_u$ and $q_d$, with $q_u+q_d=1$, such that the stock expected return
calculated with these new probabilities would equal the risk-free rate
$r$.  That is, we must demand that
\begin{equation}
S_0 (1+r)= E^Q[S_1] \equiv q_u \cdot S_1^{\rm u} + q_d \cdot S_1^{\rm d},
\label{eq:s0}
\end{equation}
where $E^Q[x]$ denotes expected value with respect to the new
probabilities $q_u$ and $q_d$.  Using the values for $S_1^{\rm u}$
and $S_1^{\rm d}$ given in Fig.~\ref{fig:binS}, we easily find that
\[
q_u = 0.3618, \quad
q_d = 0.6382.
\]
Under these probabilities, the expected value $E^Q[C_1]$ of the option
at maturity becomes $E^Q[C_1]=0.3618\times 8 + 0.6382\times 0=2.894$,
which discounted to the present time yields
$$C_0=\displaystyle\frac{E^Q[C_1]}{1+r}=\frac{2.894}{1.006}=2.88,$$
thus recovering the same price found with the delta-hedging argument.

Note that under the fictitious probability $q_u$ and $q_d$, all
financial assets (bank account, stock, and option) in our binomial
model yield exactly the same riskless rate $r$. Probabilities that
have this property of `transforming' risky assets into seemingly
risk-free ones are called an {\it equivalent martingale
measure}. Martingale measures is a topic of great relevance in
Finance, as will be discussed in more detail in Sec.~IV.

In closing this subsection, it should be noted that the one-step
binomial model considered above can be easily generalized to a
binomial tree with, say, $N$ time steps.  But for want of space this
will not be done here. (I anticipare here, however, that Black-Scholes
model to be considered later corresponds precisely to the
continuous-time limit of the binomial multistep model.) It is perhaps
also worth mentioning that binomial models are often used in practice
to price exotic derivatives, for which no closed formula exists, since
such models are rather easy to implement on the computer; see, e.g.,
\cite{hull} for more details on binomial models.

\subsection{Put-Call parity}

In the previous subsection I only considered the price of a
(European) call option, and the attentive reader might have wondered
how can one  determine the price of the corresponding put option.  It turns
out that there is a simple relationship between European put and call
options, so that from the price of one of them we can obtain the price
of the other. To see this, form the following portfolio:
i) buy one stock $S$ and one put option $P$ on this stock with
strike price $K$ and maturity $T$, and ii) short one call option
$C$ with the same strike and maturity as the put option. The value of
such portfolio would thus be
\begin{equation}
V=S+P-C. \label{eq:VPSC}
\end{equation}
Now from (\ref{eq:payoff}) and (\ref{eq:payoffP}), one immediately
sees that at expiration we have $P-C=K-S$, so that the value of the
above portfolio at time $T$ becomes simply
\begin{equation}
V_T=K.
\end{equation}
Since this portfolio has a known (i.e., riskless) value at time $t=T$,
it then follows from the no-arbitrage condition that its value at any
time $0\le t\le T$ must be given by
\begin{equation}
V=Ke^{-r(T-t)}, \label{eq:VK}
\end{equation}
where $r$ is the risk-free interest rate.  Inserting (\ref{eq:VK})
into (\ref{eq:VPSC}) immediately yields the so-called put-call parity
relation:
\begin{equation}
P=C-S+Ke^{-r(T-t)}.
\end{equation}

\section{Brownian motion and stochastic calculus}

\subsection{One-dimensional random walk}
\label{sec:1DBM}

Every physics graduate student is familiar, in one way or another,
with the concept of a Brownian motion. The customary introduction
\cite{reif} to this subject is through the notion of a random walk, in
which the anecdotal drunk walks along a line taking at every time
interval $\Delta t$ one step of size $l$, either to the right or to
the left with equal probability. The position, $X(t)$, of the walker
after a time $t=N\Delta t$, where $N$ is the number of steps taken,
represents a {\it stochastic process}. (See Appendix A for a formal
definition of random variables and stochastic processes.)  As is well
known, the probability $P(X(t)=x)$ for the walker to be found at a
given position $x=nl$, where $n$ is an integer, at given time $t$, is
described by a binomial distribution \cite{reif}.

Simply stated, the Brownian motion is the stochastic process that
results by taking the random walk to the continuous limit: $\Delta
t\to0$, $l\to0$, $N\to\infty$, $n\to\infty$ such that $t=N\Delta t$
and $x=nl$ remain finite. (A more formal definition is given below.)
Here, however, some caution with the limits must be taken to ensure
that a finite probability density $p(x,t)$ is obtained: one must take
$\Delta t\to 0$ and $l\to0$, such that $l^2=\sigma {\Delta t}$, where
$\sigma$ is a constant.  In this case one obtains that $p(x,t)$ is
given by a Gaussian distribution \cite{reif}:
\begin{equation} p(x,t) =
{1\over\sqrt{2\pi\sigma^2t}} \, \exp\left\{-{x^2\over2\sigma^2 t}\right\}.
\end{equation}

At this point let us establish some notation.  Let $X$ be a random
variable with probability density function (pdf) given by
$p(x)$. [Following standard practice, we shall denote a random
variable by capital letters, while the values it takes will be written
in small letters]. The operator for expectation value will be denoted
either as $E[\cdot]$ or $<\cdot>$, that is,
\begin{equation}
E[f(X)] \equiv \left<f(X)\right>  = \int_{-\infty}^\infty f(x)p(x)dx,
\end{equation}
where $f(x)$ is an arbitrary function.  Although the angular-bracket
notation for expectation value is preferred by physicists, we shall
often use the $E$ notation which is more convenient for our purposes.

A Gaussian or normal distribution with mean $m$ and standard deviation
$\sigma$ will be denoted by ${\cal N}(m,\sigma)$, whose pdf is
\begin{equation} 
p_{\cal N}(x,t) = {1\over\sqrt{2\pi\sigma^2}} \,
\exp\left\{-{(x-m)^2\over2\sigma^2}\right\}.
\end{equation}
  Let us also
recall that the (nonzero) moments of the Gaussian distribution are as
follows
\begin{eqnarray}
E[X]&=&m,\qquad E[X^2]=\sigma^2,\\
E[X^{2n}]&=&1\cdot3\cdot5\cdot...\cdot(2n-1) \, \sigma^{2n}. \label{eq:Ex2n}
\end{eqnarray}

\subsection{Brownian motion and white noise}

We have seen above that a 1D Brownian motion can be thought of as the
limit of a random walk after infinitely many infinitesimal
steps. This formulation was first given in 1900 by Bachelier
\cite{bachelier} who also showed the connection between Brownian
motion and the diffusion equation (five years before Einstein's famous
work on the subject \cite{einstein}). It is thus telling that the
first theory of Brownian motion was developed to model financial asset
prices! A rigorous mathematical theory for the Brownian motion was
constructed by Wiener \cite{wiener} in 1923, after which the Brownian
motion became also known as the {Wiener process}.

\begin{definition}
The standard Brownian motion or Wiener process $\{W(t), t\ge0\}$ is a
stochastic process with the following properties:
\begin{enumerate}
\item $W(0)=0$.
\item The increments $W(t)-W(s)$ are stationary and independent.
\item For $t>s$, $W(t)-W(s)$ has a normal  distribution ${\cal N}(0,\sqrt{t-s})$.
\item The trajectories are continuous (i.e., ``no jumps'').
\end{enumerate}
\end{definition}

The stationarity condition implies that the pdf of $W(t)-W(s)$, for
$t>s$, depends only on the time difference $t-s$. (For a more precise
definition of stationary processes see Appendix A.)  Now, it is not
hard to convince oneself that conditions 2 and 3 imply that $W(t)$ is
distributed according to ${\cal N}(0,\sqrt{t})$ for $t>0$. In
particular, we have $E[W(t)]=0$ for all $t\ge 0$.  Furthermore, one
can easily show that the covariance of the Brownian motion is given by
$$E[W(t)W(s)]=s,\quad {\rm for} \quad t>s.$$ It is also clear from the
definition above that the Brownian motion is a Gaussian process (see
Appendix A for the formal definition of Gaussian processes). Then, since
a Gaussian process is fully characterized by its mean and covariance,
we can  give the following alternative definition of the Brownian
motion.

\begin{definition}
The standard Brownian motion or Wiener process $\{W(t),\; t\ge0\}$ is
a Gaussian process with  $E[W(t)]=0$ and  $E[W(t)W(s)]=\min(s,t)$.
\end{definition}

The Brownian motion has the important property of having {\it bounded
quadratic variation}. To see what this means, consider
a partition $\{t_i\}_{i=0}^n$ of the interval $[0,t]$, where $0=t_{0}
< t_{1} < \ldots < t_{n}=t$.  For simplicity, let us take equally
spaced time intervals: $\displaystyle t_i-t_{i-1}=\Delta
t=\frac{t}{n}$.  The quadratic variation of $W(t)$ on $[0,t]$ is
defined as 
\begin{equation}
Q_n=\sum_{i=0}^n \Delta W_i^2,
\end{equation}
where $\Delta W_i=W(t_i)-W(t_{i-1})$. Since $\Delta W_i$ is
distributed according to ${\cal N}(0,\sqrt{\Delta t})$ we have that
$E[\Delta W^2]=\Delta t$, which implies that 
\begin{equation}
E[Q_n]=t.\label{eq:EQn}
\end{equation}
Furthermore, using the fact that the increments $\Delta W_i$ are
independent and recalling that the variance of the sum of independent
variables is the sum of the variances, we get for the variance of
$Q_n$:
\begin{eqnarray*}
\rm{var} [Q_n]&=&\sum_{i=0}^n {\rm var}[\Delta W_i^2]
=\sum_{i=0}^n \left\{E[\Delta W_i^4]-\left(E[\Delta W_i^2]\right)^2\right\}\\
&=&\sum_{i=0}^n \left[3 (\Delta t)^2- (\Delta t)^2\right]= \frac{2t^2}{n},
\end{eqnarray*}
where in the third equality we used (\ref{eq:Ex2n}) and the fact that
$\Delta W_i$ has distribution ${\cal N}(0,\sqrt{\Delta t})$. We thus see that 
\begin{equation}
\rm{var} [Q_n]\to 0, \quad \mbox{as}\quad n\to\infty.\label{eq:varQ}
\end{equation}
On the other hand, we have that 
\begin{equation}
\rm{var} [Q_n]
=E\left[\left(Q_n-E[Q_n]\right)^2\right]
=E\left[\left(Q_n-t\right)^2\right],\label{eq:varQ2}
\end{equation}
where in the last equality we have used (\ref{eq:EQn}). Comparing
(\ref{eq:varQ}) and (\ref{eq:varQ2}) then yields
$$\lim_{n\to\infty}E\left[\left(Q_n-t\right)^2\right]=0.$$

We have thus proven that {\it $Q_n$ converges to $t$ in the mean
square sense}.  This fact suggests that $\Delta W^2$ can be thought of
as being of the order of $\Delta t$, meaning that as $\Delta t\to 0$ the
quantity $\Delta W^2$ ``resembles more and more'' the deterministic
quantity $\Delta t$.  In terms of differentials, we write
\begin{equation}
[dW]^2=dt. \label{eq:dW2}
\end{equation}
Alternatively, we could say that $dW$ is of order $\sqrt{dt}$:
\begin{equation}
dW=O(\sqrt{dt}). \label{eq:dW}
\end{equation}
(I remark parenthetically that the boundedness of the quadratic
variation of the Brownian motion should be contrasted with the fact
that its total variation, $A_n=\sum_{i=0}^n \left|\Delta W_i\right|$,
is unbounded, that is, $A_n\to\infty$ as $n\to\infty$, with
probability 1; see \cite{mikosch}.)

\begin{figure}
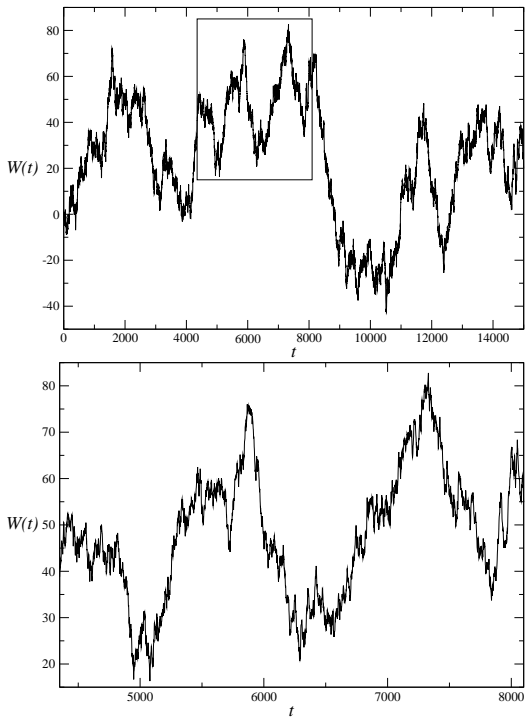

\begin{center}
\includegraphics*[width=.8\columnwidth]{fig7a.eps}
\includegraphics*[width=.8\columnwidth]{fig7b.eps}
\caption{Self-similarity of a Brownian motion path. In (a) we plot a 
path of a Brownian motion with 15000 time steps. The curve in (b) 
is a blow-up of the region delimited by a rectangle in (a), where we
have rescaled the $x$ axis by a factor 4 and the $y$ axis by a factor
2. Note that the graphs in (a) and (b) ``look the same,'' statistically
speaking. This process can be repeated indefinitely.}
\label{fig:self}
\end{center}
\end{figure}

Another important property of the Brownian motion $W(t)$ is the fact
that it is {\it self-similar} (or more exactly {\it self-affine}) in
the following sense:
\begin{equation}
W(at)\stackrel{d}{=}a^{1/2}W(t), \label{eq:a05}
\end{equation}
for all $a>0$. Here $\stackrel{d}{=}$ means equality in the sense of
probability distribution, that is, the two processes $W(at)$ and
$a^{1/2}W(t)$ have exactly the same finite-dimensional distributions
$p(x_1,t_1;...,x_n,t_n)$ for any choice of $t_i$, $i=1,...,n$, and
$n\ge 1$.  Self-similarity means that any finite portion of a Brownian
motion path when properly rescaled is (statistically) indistinguishable
from the whole path.  For example, if we `zoom in' in any given region
(no matter how small) of a Brownian motion path, by rescaling the time
axis by a factor of $a$ and the vertical axis by a factor of
$\sqrt{a}$, we obtain a curve similar (statistically speaking) to the
original path. An example of this is shown in Fig.~\ref{fig:self}. In
the language of fractals, we say that a trajectory of a Brownian
motion is a {\it fractal curve} with fractal dimension $D=2$.

The self-similarity property implies that sample paths of a Brownian
motion are nowhere differentiable (technically, with probability 1).
A formal proof of this fact, although not difficult, is beyond the
scope of the present notes, so that here we shall content ourselves
with the following heuristic argument. Suppose we try to compute the
derivative of $W(t)$ in the usual sense, that is,
$$\frac{dW}{dt}=\lim_{\Delta t\to 0}
\frac{\Delta W}{\Delta t}=\lim_{\Delta t \to 0}
\frac{W(t+\Delta t)-W(t)}{\Delta t}.$$ But since $\Delta W$ is of order 
$\sqrt{\Delta t}$, it then follows that
\begin{equation}
\frac{\Delta W}{\Delta t}=O\left(\frac{1}{\sqrt{\Delta t}}\right), \label{eq:DW}
\end{equation}
so that $dW/dt=\infty$ as $\Delta t\to0$.

Although the derivative of $W(t)$ does not exist as a regular
stochastic process, it is possible to give a mathematical meaning to
$dW/dt$ as a {\it generalized process} (in the sense of generalized
functions or distributions). In this case, the derivative of the
$W(t)$ is called the {\it white noise} process $\xi(t)$:
\begin{equation}
\xi(t)\equiv \frac{dW}{dt}.
\label{eq:xi}
\end{equation} 
I shall, of course, not attempt to give a rigorous definition of
the white noise, and so the following intuitive argument will suffice.
Since according to (\ref{eq:DW}) the derivative $\displaystyle
\frac{dW}{dt}$ diverges as $\displaystyle
\frac{1}{\sqrt{dt}}$, a simple power-counting
argument suggests that integrals of the form
\begin{equation}
I(t)=\int_0^t g(t')\xi(t')dt', \label{eq:It}
\end{equation}
should converge (in some sense); see below.

In physics, the white noise $\xi(t)$ is simply `defined' as a `rapidly
fluctuating function' \cite{reif} (in fact, a generalized stochastic
process) that satisfies the following conditions
\begin{eqnarray}
\left<\xi(t)\right>&=&0 \label{meanxi},\\
\left<\xi(t)\xi(t')\right>&=&\delta(t- t').\label{varxi}
\end{eqnarray}
These two relations give the `operational rules' from which quantities
such as the mean and the variance of the integral $I(t)$ in
(\ref{eq:It}) can be calculated.   It is convenient, however, to have
an alternative definition of stochastic integrals in terms of regular
stochastic process.  Such a construction was first given by the
Japanese mathematician It\^o \cite{ito}.

\subsection{It\^o stochastic integrals}

Using (\ref{eq:xi}), let us first rewrite integral (\ref{eq:It})
as an ``integral over the Wiener process'' $W(t)$:
\begin{equation}
I(t)=\int_0^t g(t')dW(t').
\end{equation}
The idea then is to define this integral as a kind of
Riemann-Stieltjes integral. We thus take a partition $\{t_i\}_{i=0}^n$
of the interval $[0,t]$ and consider the partial sums
\begin{equation}
I_n= \sum_{i=1}^{n}g(t_{i-1})\Delta W(t_i)\equiv
\sum_{i=1}^{n}g(t_{i-1}) [W(t_i) - W(t_{i-1})]. \label{eq:In}
\end{equation}
The function $g(t)$ above must satisfy certain appropriate conditions
\cite{mikosch}, the most important one being that $g(t)$ be a {\it
non-anticipating function}.  This means, in particular, that the value
$g(t_{i-1})$ in (\ref{eq:In}) is independent of the `next increment'
$\Delta W(t_i)$ of the Brownian motion.  [For this reason, choosing to
evaluate $g(t)$ at the beginning of the interval $\Delta
t_i=t_i-t_{i-1}$ is a crucial point in the definition of the It\^o
stochastic integral.  Another possible choice is to evaluate $g(t)$ at
the mid point $t^*=(t_{i-1}+t_i)/2$, which leads to the Stratonovich
integral \cite{oksendal}. In these notes I shall only consider It\^o
integrals.]

Under the appropriate conditions on $g(t)$, it is then possible to
show that the partial sums $I_n$ converge in the {\it mean square
sense}. That is, there exists a process $I(t)$ such that
\begin{equation}
E\left[\left(I_n-I(t)\right)^2\right]\to0 \quad{\rm as} \quad n\to\infty.
\end{equation}

Using the fact that $g(t)$ is non-anticipating and that $E\left[\Delta
W(t)\right]=0$, it follows immediately from the definition
(\ref{eq:In}) that $I(t)$ has zero mean:
\begin{equation}
E[I(t)]=E\left[\int_0^tg(t')dW(t')\right]=0, \label{meanI}
\end{equation}
It is also possible to show that stochastic integrals obey the
so-called {\it isometry property}:
\begin{equation}
E\left[\left\{I(t)\right\}^2\right]=E\left[\left(\int_0^tg(t')dW(t')\right)^2\right] =
\int_0^tE\left[g^2(t')\right]dt'.\label{varI}
\end{equation}
We now see that the true meaning of conditions (\ref{meanxi}) and
(\ref{varxi}) is given by properties (\ref{meanI}) and (\ref{varI}),
for the particular case when $g(t)$ is a
deterministic function.
 
The It\^o integral does not conform to the usual integration rules
from deterministic calculus. An example is the formula below
$$\int_0^t W dW =\frac{1}{2}W(t)^2 -\frac{1}{2}t,$$ which is left as
an exercise for the reader \cite{gardiner}. It\^o integrals offer
however a convenient way to define (and deal with) stochastic
differential equations, as we will see next.

\subsection{Stochastic differential equations}

Physicists are quite familiar with  differential equations
involving stochastic terms, such as the Langevin equation
\begin{equation}
\frac{dv}{dt}=-\gamma v + \sigma \xi(t), \label{eq:langevin}
\end{equation}
which describes the motion of a Brownian particle in a viscous liquid
\cite{reif}.  Here $\gamma$ is the viscosity of the fluid and
$\sigma$ is the `amplitude' of the fluctuating force acting on the
Brownian particle. (These parameters are usually considered to be
constant but in general they could be non-anticipating functions of
time.)  Equation (\ref{eq:langevin}) does not however make much
mathematical sense, since it evolves a quantity, namely, the
derivative $\xi(t)$ of the Brownian motion, that does not even exist
(except as a generalized process).  Nevertheless, it is possible to
put this equation on a firm mathematical basis by expressing it as a
stochastic integral equation. First we rewrite (\ref{eq:langevin}) as
\begin{equation}
dv=-\gamma v dt + \sigma dW,
\end{equation}
which upon integration yields
\begin{equation}
v(t)=v(0)-\int_0^t \gamma v(t')dt'+\int_0^t \sigma dW(t').
\end{equation}
This integral equation now makes perfectly good sense---in fact,
its solution can be found explicitly \cite{gardiner}. 
 
Let us now consider more general stochastic differential equations (SDE)
of the form
\begin{equation}
dX=a(X,t)dt + b(X,t)dW, \label{eq:dX}
\end{equation}
where $a(x,t)$ and $b(x,t)$ are known functions. Note that this
`differential equation' is actually a short-hand notation for
the following stochastic integral equation
\begin{equation}
X(t)=X(0)+\int_0^ta(X,t')dt' + \int_0^t b(X,t')dW(t'). \label{eq:X}
\end{equation}
Under certain condition on the functions $a(x,t)$ and $B(x,t)$, it is
possible to show (see, e.g., \cite{oksendal}) that the SDE
(\ref{eq:dX}) has a unique solution $X(t)$. 

Let us discuss another simple SDE, namely, the Brownian motion with
drift:
\begin{equation}
dX=\mu dt + \sigma dW, \label{eq:BMD}
\end{equation}
where the constant $\mu$ represents the mean drift velocity.
Integrating (\ref{eq:BMD}) immediately yields the process
\begin{equation}
X(t)=\mu t + W(t),
\end{equation}
whose pdf is 
\begin{equation}
p(x,t)=\frac{1}{2\pi\sigma^2 t} \exp\left\{\frac{(x-\mu t)^2}{2\sigma^2 t}\right\}. \label{eq:pdfBMD}
\end{equation}
Another important example of a (linear) SDE that can be solved
explicitly is the geometric Brownian motion that will be discussed
shortly. But before doing that, let us discuss a rather useful result known
as {\it It\^o lemma} or {\it It\^o formula}.

\subsection{It\^o formula}

Consider the generic process $X(t)$ described by the SDE (\ref{eq:dX}),
and suppose  that we have a new stochastic process $Z$ defined  by 
\begin{equation}
Z(t)=F(X(t),t), 
\end{equation}
for some given function $F(x,t)$. We now wish to find the local
dynamics followed by the $Z(t)$, that is, the SDE whose solutions
corresponds to the process $Z(t)$ above. The answer is given by the
It\^o formula that we now proceed to derive.

First, consider the Taylor expansion of the function $F(X,t)$:
\begin{eqnarray}
dF &= & {\partial F\over\partial t}dt+{\partial F\over\partial
x}dX+{1\over 2}{\partial^2 F\over\partial x^2}(dX)^2+\cr&&+\; {1\over
2}{\partial^2 F\over\partial t^2}(dt)^2 + {1\over 2}{\partial^2
F\over \partial t\partial x}dtdX + ...\label{eq:dZ1}
\end{eqnarray}
Note, however, that
\begin{eqnarray}
(dX)^2 &=&  b^2dW^2+ 2 a b \,  dtdW + a^2(dt)^2 \cr
&=& b^2 dt + O(dt^{3/2}), \label{eq:dXsq}
\end{eqnarray}
where we used the fact that $dW^2=dt$ and $dtdW = O(dt^{3/2})$. (Here
we have momentarily omitted the arguments of the functions $a$ and $b$
for ease of notation.)  Inserting (\ref{eq:dXsq}) into (\ref{eq:dZ1})
and retaining only terms up to order $dt$, we obtain
\begin{eqnarray}
dF &=& \left[{\partial F\over\partial t}+{1\over 2}b^2\, {\partial^2
F\over\partial x^2}\right] dt + b\, {\partial F\over\partial
x}dX,
\label{eq:dZZ}
\end{eqnarray}
which is known as It\^o formula.  Upon using (\ref{eq:dX}) in the
equation above, we obtain It\^o formula in a more explicit fashion
\begin{eqnarray}
dF &=& \left[{\partial F\over\partial t}+a(X,t){\partial
F\over\partial x}+{1\over 2}b^2(X,t){\partial^2 F\over\partial
x^2}\right] dt\cr +&& b(X,t){\partial F\over\partial x}dW,
\label{eq:dZ}
\end{eqnarray}
What is noteworthy about this formula is the fact that the fluctuating
part of the primary process $X(t)$ contributes to the drift of the
derived process $Z(t)=F(t,X)$ through the term ${1\over
2}b^2(t,X){\partial^2 F\over\partial x^2}$. We shall next use It\^o
formula to solve explicitly a certain class of linear SDE's.

\subsection{Geometric Brownian motion}

\label{sec:GMB}

A stochastic process of great importance in Finance is the so-called
geometric Brownian notion, which is defined as the solution to the
following SDE
\begin{equation}
dS=\mu S dt + \sigma S dW, \label{eq:geoMB}
\end{equation}
where $\mu$ and $\sigma$ are constants, subjected to a generic initial
condition $S(t_0)=S_0$. Let us now perform the following change of
variables $Z=\ln S$. Applying It\^o formula (\ref{eq:dZ}) with $a=\mu
S$, $b=\sigma S$ and $F(S)=\ln S$, it then follows that
\begin{equation}
dZ = \left(\mu - {1\over 2} \sigma^2\right) dt +\sigma dW,
\end{equation}
which upon integration yields
\begin{equation}
Z(t) =Z_0+ \left(\mu - {1\over 2} \sigma^2\right) (t-t_0) + \sigma [W(t)-W(t_0)],\label{eq:Zt}
\end{equation}
where  $Z_0=\ln S_0$. Reverting to the variable $S$ we obtain
the explicit solution of the SDE (\ref{eq:geoMB}):
\begin{equation}
S(t)=S_0\exp\left\{\left(\mu - {1\over 2} \sigma^2\right) (t-t_0) +
\sigma [W(t)-W(t_0)]\right\}.
\label{eq:geoMBS}
\end{equation}

From (\ref{eq:Zt}) we immediately see that $Z(t)-Z_0$ is distributed
according to ${\cal N}\left(\left(\mu -\frac{1}{2}
\sigma^2\right)\tau,\sigma\sqrt{\tau}\right)$, where $\tau=t-t_0$.  It then follows that
the geometric Brownian motion with initial value $S(t_0)=S_0$ has the
following log-normal distribution:
\begin{equation}
p(S,t;S_0,t_0)=\frac{1}{\sqrt{2\sigma^2 \tau} S}
\exp\left\{-\frac{\left[\ln\left(\frac{S}{S_0}\right)- (\mu - {1\over
2} \sigma^2)\tau\right]^2}{2\sigma^2\tau}\right\}. \label{eq:logN}
\end{equation}  
The geometric Brownian motion is the basic model for stock price
dynamics in the Black-Scholes framework, to which we now turn.

\section{The Standard Model of Finance}

\label{sec:BS}

\subsection{Portfolio dynamics and arbitrage}

Consider a financial market with only two assets: 
a risk-free bank account $B$ and a stock $S$.  In vector
notation, we write $\vec{S}(t)=(B(t),S(t))$ for the asset {\it
price vector} at time $t$.  A {\it portfolio} in this market consists
of having an amount $x_0$ in the bank and owing $x_1$ stocks. The
vector $\vec{x}(t)=(x_0(t),x_1(t))$ thus describes the time evolution
of your portfolio in the $(B,S)$ space. Note that $x_i<0$ means a
short position on the $i$th asset, i.e., you `owe the market' $|x_i|$
units of the $i$th asset.  Let us denote by $V_{\vec{x}}(t)$ the money
value of the portfolio $\vec{x}(t)$:
\begin{equation}
V_{\vec{x}}=\vec{x}\cdot\vec{S}=x_0 B + x_1 S, \label{eq:V}
\end{equation}
where the time dependence has been omitted for clarity. We shall also
often suppress the subscript from $V_{\vec{x}}(t)$ when there is no
risk of confusion about to which portfolio we are referring.

A portfolio is called {\it self-financing} if no money is taken from
it for `consumption' and no additional money is invested in it, so
that any change in the portfolio value comes solely from changes in the asset
prices.  More precisely, a portfolio $\vec{x}$ is self-financing if
its dynamics is given by
\begin{equation}
dV_{\vec{x}}(t) = \vec{x}(t)\cdot d\vec{S}(t), \quad t\ge 0. \label{eq:dV}
\end{equation}
The reason for this definition is that in the discrete-time case,
i.e., $t=t_n$, $n=0,1,2,...$, the increase in wealth, $\Delta V(t_n) =
V(t_{n+1})-V(t_n)$, of a self-financing portfolio over the time
interval $t_{n+1}-t_n$ is given by
\begin{equation}
 \Delta V(t_n) = \vec{x}(t_n)\cdot \Delta\vec{S}(t_n), 
\end{equation}
where $\Delta\vec{S}(t_n)\equiv\vec{S}(t_{n+1})-\vec{S}(t_n)$. This
means that over the  time interval $t_{n+1}-t_n$ the value of the
portfolio varies only owing to the changes in the asset prices
themselves, and then at time $t_{n+1}$ re-allocate the assets
within the portfolio for the next time period. Equation (\ref{eq:dV})
generalizes this idea for the continuous-time limit.  If furthermore
we decide on the make up of the portfolio by looking only at the
current prices and not on past times, i.e., if
\[
\vec{x}(t)=\vec{x}(t,\vec{S}(t)),
\]
then the portfolio is said to be {\it Markovian}. Here we shall
deal exclusively with Markovian portfolios.

As we have seen already in Sec.~\ref{sec:HSA}, an arbitrage represents
the possibility of making a riskless profit with no initial commitment
of money. A more formal definition of arbitrage is as follows.

\begin{definition}
An arbitrage is a portfolio whose value $V(t)$ obeys the following
conditions
\begin{enumerate}
\item[(i)] $V(0)=0$
\item[(ii)] $V(t)\ge 0$ with probability 1 for all $t>0$
\item[(iii)] $V(T)>0$ with {\it positive} probability for some $T>0$.
\end{enumerate}
\end{definition}

\vspace{0.5cm}

The meaning of the first condition is self-evident. The second
condition says that there is no chance of losing money, while the
third one states that there is a possibility that the portfolio will
acquire a positive value at some time $T$. Thus, if you hold this
portfolio until this arbitrage time there is a real chance that you
will make a riskless profit out of nothing. [If $P(V(T)>0)=1$ we have
a {\it strong arbitrage} opportunity, in which case we are {\it sure}
to make a profit.] As we have already discussed in Sec.~\ref{sec:HSA},
arbitrage opportunities are very rare and can last only for a very
short time (typically, of the order of seconds or a few minutes at
most).  In fact, in the famous Black-Scholes model that we will now
discuss it is assumed that there is no arbitrage at all.

\subsection{The Black-Scholes model for option pricing}

The two main assumptions of the Black-Scholes model are:

\begin{enumerate}
\item[(i)] There are two assets in the market, a bank account $B$ 
and a stock $S$, whose price dynamics are governed by the following
differential equations
\begin{eqnarray}
dB&=&rB dt, \label{eq:dB'}\\
dS&=&\mu S dt + \sigma S dW \label{eq:dS'},
\end{eqnarray}
where $r$ is the risk-free interest rate, $\mu>0$ is the stock {\it
mean rate of return}, $\sigma>0$ is the {\it volatility}, and $W(t)$
is the standard Brownian motion or Wiener process.

\item[(ii)] The  market is free of arbitrage.

\end{enumerate}

Besides these two crucial hypothesis, there are additional simplifying
(technical) assumptions, such as: (iii) there is a liquid market for
the underlying asset $S$ as well as for the derivative one wishes to
price, (iv) there are no transaction costs (i.e., no bid-ask spread),
and (v) unlimited short selling is allowed for an unlimited period of
time.  It is implied by (\ref{eq:dB'}) that there is no interest-rate
spread either, that is, money is borrowed and lent at the same rate
$r$. Equation (\ref{eq:dS'}) also implies that the stock pays no
dividend. [This last assumption can be relaxed to allow for dividend
payments at a known (i.e., deterministic) rate; see, e.g.,
\cite{bjork} for details.]

We shall next describe how derivatives can be `rationally' priced in
the Black-Scholes model. We consider first a European call option for
which a closed formula can be found. (More general European contingent
claims will be briefly considered at the end of the Section.)  Let us
then denote by $C(S,t;K,T)$ the present value of a European call
option with strike price $K$ and expiration date $T$ on the underlying
stock $S$. For ease of notation we shall drop the parameters $K$ and
$T$ and simply write $C(S,t)$.  For later use, we note here that
according to It\^o formula (\ref{eq:dZ}), with $a=\mu S$ and $b=\sigma
S$, the option price $C$ obeys the following dynamics
\begin{equation}
dC = \left[ {\partial C\over \partial t} + \mu S{\partial C\over
\partial S}+ {1\over 2}\sigma^2 S^2{\partial^2 C\over \partial
S^2}\right] dt + \sigma S{\partial C\over \partial S}dW.
\label{eq:dC}
\end{equation}

In what follows, we will arrive at a partial differential equation,
the so-called Black-Scholes equation (BSE), for the option price
$C(S,t)$.  For pedagogical reasons, we will present two alternative
derivations of the BSE using two distinct but related arguments: i)
the $\Delta$-hedging portfolio and ii) the replicating portfolio.

\subsubsection{The delta-hedging portfolio}

As in the binomial model of Sec.~\ref{sec:bin}, we consider the
self-financing $\Delta$-hedging portfolio, consisting of a long
position on the option and a short position on $\Delta$ stocks. The
value $\Pi(t)$ of this portfolio is 
\[
\Pi(t)=C(S,t)-\Delta \; S .
\]
Since the portfolio is self-financing, it follows from (\ref{eq:dV})
that $\Pi$ obeys the following dynamics
\begin{equation}
d\Pi=dC-\Delta \; dS, \label{eq:dPi}
\end{equation}
which in view of (\ref{eq:dS'}) and (\ref{eq:dC}) becomes 
\begin{eqnarray}
d\Pi &=& \left[ {\partial C\over \partial t} + \mu S{\partial C\over
\partial S}+ {1\over 2}\sigma^2 S^2{\partial^2 C\over \partial S^2} -
\mu \Delta S\right] dt\cr &&+ \; \sigma S\left({\partial C\over \partial S}-
\Delta\right)dW.
\label{eq:5}
\end{eqnarray}

We can now eliminate  the risk [i.e., the stochastic term
containing $dW$] from this portfolio by choosing
\begin{equation}
\Delta = {\partial C\over \partial S}. \label{eq:D}
\end{equation}
Inserting this back into (\ref{eq:5}), we then find
\begin{equation}
d\Pi = \left[{\partial C\over \partial t} + {1\over 2}\sigma^2
S^2{\partial^2 C\over \partial S^2}\right]dt.
\label{eq:5a}
\end{equation}
Since we now have a risk-free (i.e., purely deterministic) portfolio,
it must yield the same rate of return as the bank account, which means that
\begin{equation}
d\Pi = r\Pi dt.\label{eq:dV2}
\end{equation}
Comparing (\ref{eq:5a}) with (\ref{eq:dV2}) and using (\ref{eq:dPi}) and
(\ref{eq:D}), we then obtain the Black-Scholes equation:
\begin{equation}
{\partial C\over \partial t} + {1\over 2}\sigma^2 S^2{\partial^2 C\over \partial S^2} +  rS{\partial C\over \partial S} - rC = 0,
\label{eq:BS}
\end{equation}
which must  be solved subjected to the following boundary condition
\begin{equation}
C(S,T) = \mbox{max}(S- K, 0).
\label{eq:CTS}
\end{equation}
The solution to the above boundary-value problem can be found
explicitly (see below), but before going into that it is instructive
to consider an alternative derivation of the BSE.  [Note that the
above derivation of the BSE remains valid also in the case that $r$,
$\mu$, and, $\sigma$ are deterministic functions of time, although a
solution in closed form is no longer possible.]

\subsubsection{The replicating portfolio}

\label{sec:repli}

Here we will show that it is possible to form a portfolio on the
$(B,S)$ market that {\it replicates} the option $C(S,t)$, and in the
process of doing so we will arrive again at the BSE. Suppose then that
there is indeed a self-financing portfolio $\vec{x}(t)=(x(t),y(t))$,
whose value $Z(t)$ equals the option price $C(S,t)$ for all time $t\le
T$:
\begin{equation}
Z\equiv x B + y S = C, \label{eq:Z}  
\end{equation}
where we have omitted the time-dependence for brevity. Since the
portfolio is self-financing it follows that
\begin{equation}
dZ=x dB + y dS = (r x B+ \mu y S)dt+\sigma y S dW. \label{eq:dV3}
\end{equation}
But by assumption we have $Z=C$ and so  $dZ=dC$.  Comparing
(\ref{eq:dV3}) with (\ref{eq:dC}) and equating the coefficients
separately in both $dW$ and $dt$, we obtain
\begin{eqnarray}
&&y=\frac{\partial C}{\partial S},\label{eq:y} \\
&&{\partial C\over \partial t} -r x B + {1\over 2}\sigma^2 S^2{\partial^2 C\over \partial
S^2}=0.  \label{eq:XX}
\end{eqnarray}
Now from (\ref{eq:Z}) and (\ref{eq:y}) we get that
\begin{equation}
x=\frac{1}{B}\left[C-S\frac{\partial C}{\partial S}\right], \label{eq:x}
\end{equation}
which inserted into (\ref{eq:XX}) yields again the BSE
 (\ref{eq:BS}), as the reader can easily verify. 

We have thus proven, by direct construction, that the option $C$ can
be replicated in the $(B,S)$-market by the portfolio $(x,y)$, where
$x$ and $y$ are given in (\ref{eq:x}) and (\ref{eq:y}), respectively,
with option price $C$ being the solution of the BSE (with the
corresponding boundary condition).  [To complete the proof, we must
also show that the initial price $C_0=C(S,0)$ is the `correct' one, in
the sense that if the option price were $C_0'\ne C_0$, then there
would be an arbitrage opportunity. In fact, if $C_0'>C_0$  an
arbitrageur should short the option and invest in the replicating
portfolio, whereas if $C_0'<C_0$ he should do the opposite.]

\subsection{The Black-Scholes formula}
\label{sec:fBS}

Here we will solve equation (\ref{eq:BS}) subjected
to the boundary condition (\ref{eq:CTS}).  Following the original work
of Black and Scholes \cite{BS}, the idea is to perform a change of
variables so as to turn the BSE  into the heat equation,
which we know how to solve.  Here we will not use the original
transformation employed by these authors but a related one
\cite{wilmott}, as shown below:
\begin{equation}
\tau =  \frac{T-t}{{2/\sigma^2}}  \; , \quad
x =  \ln\left(S\over K\right), \label{eq:xt}
\end{equation}
\begin{equation}
 u(x,\tau) = e^{\alpha x +\beta^2 \tau}\; \frac{C(S,t)}{K},
\end{equation}
where
\begin{equation}
\alpha = {1\over2}\left({2r\over\sigma^2}-1\right), \ \ \beta={1\over2} 
\left({2r\over\sigma^2}+1\right) .
\end{equation}

After a somewhat tedious but straightforward algebra \cite{wilmott},
one obtains that in the new variables equation (\ref{eq:BS}) reads
\begin{equation}
{\partial u\over\partial \tau} = {\partial^2 u\over\partial x^2},
\label{eq:8}
\end{equation}
while the terminal condition (\ref{eq:CTS}) becomes an initial condition
\begin{equation}
u(x,0)=u_0(x) = \max\left(e^{\beta x} - e^{\alpha x}, 0\right) .
\label{eq:9}
\end{equation}

We now recall that the Green's function for the heat equation is
\[
G(x,x')={1\over\sqrt{4\pi\tau}}\, e^{-(x-x')^2/4\tau}, \label{eq:G}
\]
so that its generic solution for an arbitrary initial condition
$u_0(x)$ is given by
\begin{eqnarray}
u(x,\tau)&=&\int_{-\infty}^\infty u_0(x')G(x,x')dx'\cr
&=&{1\over\sqrt{4\pi\tau}}\int_{-\infty}^\infty
u_0(x')e^{-(x-x')^2/4\tau} dx'. \label{eq:u}
\end{eqnarray}
Inserting  (\ref{eq:9}) into the integral above we obtain
\begin{eqnarray}
u(\tau,x)&=&{1\over\sqrt{4\pi\tau}}\int_0^{\infty} \left(e^{\beta x'} - e^{\alpha x'}\right) e^{-(x-x')^2/4\tau} dx'\cr\cr
&=&I(\beta)-I(\alpha), \label{eq:u2}
\end{eqnarray}
where
\begin{equation}
I(a)\equiv{1\over\sqrt{4\pi\tau}}\int_0^{\infty} e^{a x'} e^{-(x-x')^2/4\tau} dx'.
\end{equation}
After completing the squares and performing some simplification, we
find that
\begin{equation}
I(a)=e^{a
x+a^2 \tau} N(d_a), \label{eq:Ia}
\end{equation}
where
\begin{equation}
d_a=\frac{x+2a\tau}{\sqrt{2\tau}},
\end{equation}
and $N(x)$ denotes the cumulative distribution function for a normal
variable ${\cal N}(0,1)$:
\begin{equation}
N(x) = {1\over\sqrt{2\pi}}\int_{-\infty}^x e^{-s^2/2} ds.
\label{eq:Nx}
\end{equation}
Inserting (\ref{eq:Ia}) into (\ref{eq:u2}) and reverting back to the
original dimensional variables, we obtain the famous Black-Scholes
formula for the price of a European call option:
\begin{equation}
C(S,t) = S N(d_1) - K e^{-r(T-t)}N(d_2),
\label{eq:fBS}
\end{equation}
where
\begin{eqnarray}
d_1 &=& \frac{\ln\left(S\over K\right) +
\left(r+ {1\over2}\sigma^2\right) \left(T-t\right)}{\sigma\sqrt{T-t}},\\ \\ d_2
&=& \frac{\ln\left(S\over K\right) +
\left(r- {1\over2}\sigma^2\right) \left(T-t\right)}{\sigma\sqrt{T-t}}.
\end{eqnarray} 
This formula is so often used in practice that it is already
pre-defined in many software packages (e.g., Excel, Matlab, Maple,
etc) as well as in most modern hand calculators with financial
functions. It should noted, however, that many people (academics and
practitioners alike) believe that the Black-Scholes model is too
idealized to describe real market situations; see Secs.~V and VII
for a brief discussion of possible extensions of the BS model.

\subsection{Completeness in the Black-Scholes model}

We have seen above that it is possible to replicate a European call
option $C(S,t)$ using an appropriate self-financing portfolio in the
$(B,S)$ market. Looking back at the argument given in
Sec.~\ref{sec:repli}, we see that we never actually made use of the
fact that the derivative in question was a call option---the nature of
the derivative appeared only through the boundary condition
(\ref{eq:CTS}). Thus, the derivation of the BSE presented there must
hold for {\it any} contingent claim!

To state this fact more precisely, let $F(S,t)$ represent the price of
an arbitrary European contingent claim with payoff $F(S,T)=\Phi(S)$,
where $\Phi$ is a known function. Retracing the steps outlined in
Sec.~\ref{sec:repli}, we immediately conclude that the price $F(S,t)$
will be the solution to the following boundary-value problem
\begin{eqnarray}
{\partial F\over \partial t} + {1\over 2}\sigma^2 S^2{\partial^2
F\over \partial S^2} + rS{\partial F\over \partial S} - rF = 0&&,
\label{eq:FBS}\\  F(S,T)=\Phi(S)&&.
\end{eqnarray}
Furthermore, if we repeat the arguments of preceding subsection and
transform the Black-Scholes equation (\ref{eq:FBS}) into the heat
equation, we obtain that $F(S,t)$ will be given by
\begin{equation}
F(S,t)={1\over\sqrt{4\pi\tau}}\int_{-\infty}^\infty \Phi(x')e^{-(x-x')^2/4\tau} dx', \label{eq:FTS'}
\end{equation}
where $\Phi(x)$ denotes the payoff function in terms of the
dimensionless variable $x$; see (\ref{eq:xt}).  Expressing this result
in terms of the original variables $S$ and $t$ yields a
generalized Black-Scholes  formula
\begin{equation}
F(S,t)= {e^{-r(T-t)}\over\sqrt{2\pi\sigma^2(T-t)}}\int_{0}^\infty
\Phi(S')e^{[\ln\left(\frac{S'}{S}\right)-(r-{1\over2}\sigma^2)(T-t)]^2}
\frac{dS'}{S'}. \label{eq:FTS}
\end{equation}

In summary, we have shown above that the Black-Scholes model is
complete. A market is said to be {\it complete} if every contingent
claim can be replicated with a self-financing portfolio on the primary
assets.  Our `proof of completeness' given above is, of course, valid
only for the case of European contingent claims with a simple payoff
function $\Phi(S)$; it does not cover, for instance, path-dependent
derivatives. It is possible however to give a formal proof that
arbitrage-free models, such as the Black-Scholes model, are indeed
complete; see Sec.~\ref{sec:symm}.

Comparing the generalized Black-Scholes formula (\ref{eq:FTS}) with
the pdf of the geometric Brownian motion given in (\ref{eq:logN}), we
see that the former can be written in a convenient way as
\begin{equation}
F(S,t) = e^{-r(T-t)} E_{t,S}^Q[\Phi(S_T)],
\label{eq:fFTS}
\end{equation}
where $E_{t,S}^Q[\cdot]$ denotes expectation value with respect to
the probability density of a geometric Brownian motion with $\mu=r$,
initial time $t$, final time $T$, and initial value $S$; see (\ref{eq:logN}).
In other words, the present value of a contingent claim can be
computed simply as its discounted expected value at maturity, under an
appropriate probability measure. This idea will become more clear
after we discuss the notion of an equivalent martingale measure.

\section{Efficient markets: the martingale approach}

\subsection{Martingales}

The concept of a martingale plays a important r\^ole in finance
\cite{malliaris}.  Unfortunately, a proper introduction to martingales
requires some knowledge of probability measure theory
\cite{billingsley}. Here however we will give a rather intuitive
discussion of martingales.  For completeness we have listed in
Appendix A some basic concepts from probability theory that would be
required to make the following discussion more rigorous.

We begin by recalling that  a {\it probability space} is a triple
$(\Omega,{\cal F}, P)$, where
\begin{itemize}
\item $\Omega$ is the space of elementary events or {\it outcomes}
$\omega$. 
\item $\cal F$ is a properly chosen family of subsets of $\Omega$, 
(i.e., a $\sigma$-algebra on $\Omega$). 
\item $P$ is a   probability measure on $\cal F$.
\end{itemize}

In Finance, an outcome $\omega$ is a `market situation.' The family of
subsets $\cal F$ specifies the class of events to which probabilities
can be assigned. This is done through the concept of a
$\sigma$-algebra, whose formal definition is given in Appendix A.  A
probability measure $P$ on $\cal F$ is simply a function $P:{\cal
F}\rightarrow[0,1]$, satisfying a few `obvious requirements':
$P(\emptyset)=0$, $P(\Omega)=1$, and $P(A_1\cup A_2)=P(A_1)+P(A_2)$ if
$A_1\cap A_2=\emptyset$. An element $A$ of $\cal F$, $A\in\cal F$, is
called a ``measurable set'' or ``observable event,'' meaning that it
is possible to assign a ``probability of occurrence,'' $P(A)\in[0,1]$,
for this event. Hence, $\cal F$ is said to be the set of
`observable events.'

Suppose now that we have a random function $X:\Omega\to{\bf R}$.  If
to every subset of $\Omega$ of the form $\{\omega:a \le X(\omega) \le
b\}$ there corresponds an event $A\subset{\cal F}$, then the function
$X$ is said to be {\it measurable} with respect to $\cal F$ or simply
$\cal F$-{\it measurable}.  What this means is that it is
possible to `measure' (i.e., assign a probability to) events of the
form $\{a\le X\le b\}$ through the obvious definition: $P(\{a\le X\le
b\})\equiv p(A)$. A $\cal F$-measurable function $X$ is called a {\it
random variable}.

Let us next consider the notion of an ``information flow.''  In a
somewhat abstract way, we will represent the {\it information}
available to an observer up to time $t$ as a $\sigma$-algebra ${\cal
F}_t\subset{\cal F}$.  In the context of Finance, the information set
${\cal F}_t$ would contain, for instance, the price history
up to time $t$ of  all assets in the economy. It is
natural to assume that ${\cal F}_s\subset{\cal F}_t$ for $s\le t$,
since we expect that, as time goes on, we gain new information (and do
not discard the old ones).  Such a collection of $\sigma$-algebras
represents an ``information flow,'' or, more technically, a {\it
filtration}.

\begin{definition}
A filtration or information flow is a collection $\{{\cal
F}_t\}_{t\ge0}$ of $\sigma$-algebras ${\cal F}_t\subset {\cal F}$ such
that $${\cal F}_s\subset{\cal F}_t, \quad {\rm for}\quad 0\le s\le
t.$$
\end{definition}

Suppose now we have a stochastic process $X_t$ defined on
$(\Omega,{\cal F},P)$.  (Think of $X_t$ as being, say, the price of a
given stock.) If the values of $X_t$ can be completely determined from
the information ${\cal F}_t$, for all $t\ge0$, then the process $X_t$
is said to be {\it adapted} to the filtration $\{{\cal F}_t\}_{t\ge
0}$.

\begin{definition}
The process $X_t$ is {\it adapted} to the filtration $\{{\cal
F}_t\}_{t\ge 0}$ if $X_t$ is ${\cal
F}_t$-measurable for all $t\ge0$.
\end{definition}

A stochastic process $X_t$ naturally generates an information flow,
denoted by ${\cal F}^X_t$, which represents the ``information
contained in the trajectories of $X(t)$ up to time $t$.'' A process
$X(t)$ is obviously adapted to its natural filtration ${\cal F}^X_t$.

A last piece of mathematics is necessary to define a martingale,
namely, the notion of conditional expectation.  This appears in
connection with the crucial question of
how the information, ${\cal F}_{t_0}$, available at present time
influences our knowledge about future values of $X(t)$.  If $X(t)$ and
${\cal F}_{t_0}$ are not independent, then it is reasonable to expect that
the information available up to the present time reduces the
uncertainty about the future values of $X(t)$. To reflect this gain of
information is useful to introduce a new stochastic process
$$Z(t)=E[X_t|{\cal F}_{t_0}], \quad t>t_0,$$ where the symbol
$E[X_t|{\cal F}_{t_0}]$ represents ``the expected value of $X_t$,
contingent on the information gathered up to time $t_0$.'' 

A precise definition of conditional expectation is beyond the scope of
these notes.  Here it will suffice to say that given a random variable
$Y$ on a probability space $(\Omega,{\cal F}, P)$ and another
$\sigma$-algebra ${\cal F}'\subset {\cal F}$, it is possible to define
a random variable $Z=E[Y|{\cal F}']$, which represents ``the expected
value of $Y$, given the information contained in ${\cal F}'$.'' The
variable $Z$ is a coarser version of the original variable $Y$, in the
sense that we have used the information on ${\cal F}'$ to reduce the
uncertainty about $Y$.  The following two properties of conditional
expectations will be necessary later:
\begin{eqnarray}
 &&E[Y|{\cal F}_{t}]=Y,  \quad \mbox{ if $Y$ is ${\cal F}_{t}$-measurable}.
\label{eq:EY}\\
&&E[E[Y|{\cal F}_{t}]]=E[Y].\label{eq:EEY}
\end{eqnarray}
The first property above is somewhat obvious: if $Y$ is ${\cal
F}_{t}$-measurable then $Y$ and ${\cal F}_{t}$ `contain the same
information,' hence taking expectation of $Y$ conditional to ${\cal
F}_{t}$ does not reduce the uncertainty about $Y$. The second property
is the so-called {\it law of iterated expectations}, and basically
represents the law of total probability.

After these mathematical preliminaries, we are now in a position to
define martingales.

\begin{definition}
A stochastic process $M_t$ is called a martingale with respect to the
filtration $\{{\cal F}_t\}_{t\ge 0}$ if 
\begin{enumerate}
\item[(i)] $M_t$ is adapted to the filtration  $\{{\cal F}_t\}_{t\ge 0}$
\item[(ii)] $E[|M_t|]<\infty$ for all $t\ge0$
\item[(iii)] $E[M_t|{\cal F}_{t_0}]=M_{t_0}$ for all $t\ge t_0$
\end{enumerate}
\end{definition}

Condition (i) simply says that $M_t$ can be determined from the
information available up to time $t$, whereas condition (ii) is a
technicality. The defining property of a martingale is therefore
condition (iii), which is usually referred to as the {\it martingale
condition}. It says that the best prediction of future values of the
process $M(t)$, contingent on the information available at the present
time, is the current value $M_{t_0}$.

Because of property (iii), a martingale is usually described as a
``fair game.''  To see why this is so, suppose that $M_n$ represents
the fortune of a gambler at time $n$. (For convenience let us assume
here that time is discrete.) The difference $h_n=M_n-M_{n-1}$ is then
the amount the gambler wins on the $n$th play (a negative win is of
course a loss). Now let us compute the gambler's expected gain on the
$(n+1)$th play, given the information up to time $n$:
\begin{eqnarray}
E[h_{n+1}|{\cal
F}_n]&=&E[M_{n+1}-M_n|{\cal F}_n]\cr&=&E[M_{n+1}|{\cal F}_n] -E[M_n|{\cal
F}_n]\cr&=&M_n-M_n\cr&=&0,
\end{eqnarray}
where in the third equality we used the martingale property and
rule (\ref{eq:EY}).  We thus have that at each new play of the game the
expected gain is null, and in this sense it is a ``fair'' game.

A Brownian motion $W(t)$ is a martingale with respect to its natural
filtration.  To show this, we only need to verify  the martingale
condition (since the other two conditions are trivially fulfilled):
\begin{eqnarray}
E[W(t)|{\cal F}_{t_0}]&=&E[W(t)-W(t_0)+W(t_0)|{\cal
F}_{t_0}]\cr&=&E[W(t)-W(t_0)|{\cal F}_{t_0}]+E[W(t_0)|{\cal
F}_{t_0}]\cr &=&0+W(t_0)\cr &=&W(t_0).
\end{eqnarray}
In the third equality above we used the fact that the increments
$W(t)-W(t_0)$ are independent of ${\cal F}_{t_0}$ and have zero mean,
together with property (\ref{eq:EY}). It is also possible to show that
It\^o stochastic integrals are martingales. Indeed, the theory of
stochastic integration is intimately connected with martingale theory
\cite{shiryaev}.

Another important property of martingales is that their expected value
remains constant in time: $$E[M_0]=E[E[M_t|{\cal F}_0]]=E[M_t],$$
where in first equality we used the martingale property, while in the
second equality property (\ref{eq:EEY}) was used. Thus, a necessary (but not
sufficient) condition for a process to be a martingale is that it have
no drift.  Therefore, a diffusion process of the form (\ref{eq:dX}) is
not a martingales unless the drift term vanishes.  For this reason, the
geometric Brownian motion (\ref{eq:geoMB}) is not a martingale. It is
possible, however, to introduce a new probability measure $Q$, with
respect to which the geometric Brownian motion becomes a standard
Brownian motion and hence a martingale, as discussed next.

\subsection{Equivalent martingale measures}

Recall that a probability measure $P$ on a measurable space
$(\Omega,{\cal F})$ is a function $P:{\cal F}\to[0,1]$ that assigns to
every event $A\subset{\cal F}$ a real number $P(A)\in[0,1]$.  Suppose
now we have another probability measure $Q$ defined on the same space
$(\Omega,{\cal F})$. We say that the probability measures $P$ and $Q$
are equivalent if the following condition is satisfied:
\begin{equation}
Q(A)=0\Longleftrightarrow P(A)=0, \quad \mbox{for all}\quad  A\in{\cal F}.\label{eq:QP}
\end{equation}
To get a better grasp on the meaning of the condition above, consider
a random variable $X$ [on $(\Omega,{\cal F},P)$]. If $Q$ is a
probability measure equivalent to $P$, then condition (\ref{eq:QP})
implies that there exists a function $\rho(X)$ such that 
expected values w.r.t~$Q$ are calculated in the following way
\begin{equation}
E_Q[g(X)]=E_P[\rho(X) g(X)], \label{eq:EQ}
\end{equation}
where $g(x)$ is an arbitrary function.  Alternatively, we may write
(\ref{eq:EQ}) in terms of probability densities: 
\begin{equation}
f_Q(x)=\rho(x)f_P(x),\label{eq:fQP}
\end{equation}
where $f_P(x)$ denotes the probability density of $X$ w.r.t~the
measure $P$ and $f_Q(x)$ is the density w.r.t~$Q$.  (The function
$\rho(x)$ is called the Radon-Nikodym derivative of measure $Q$ with
respect to measure $P$ \cite{billingsley}.)

Consider now the Brownian motion with drift
\begin{equation}
\tilde{W}(t)=a t + W(t), \quad 0\le t \le T \label{eq:Wtil}
\end{equation}
where $a$ is some constant. (The finite-horizon condition $t<T$ is a
technicality that is not relevant for our purposes.)  As already
noted, $\tilde{W}(t)$ is not a martingale (since its expected value is
not constant). However, there is an equivalent probability measure
$Q$, with respect to which the process $\tilde{W}(t)$ becomes the
standard Brownian motion (and hence a martingale). This result  is
known as Girsanov theorem.

\begin{thm}[Girsanov theorem]
The process $\tilde{W}(t)$ given in (\ref{eq:Wtil}) is a standard
Brownian motion with respect to the probability measure $Q$
defined by
\begin{equation}
f_Q(\tilde{x},t)=M_t(\tilde{x}) f_P(\tilde{x},t),\label{eq:fQ}
\end{equation}
where $M_t$ is the process
\begin{equation}
M_t=\exp\left\{-a W_t-\frac{1}{2}a^2 t\right\}=\exp\left\{-a \tilde{W}_t+\frac{1}{2}a^2 t\right\}.
\label{eq:Mt}
\end{equation} 
\end{thm}

\medskip

\noindent{\it Proof}. For a formal proof see, e.g., \cite{oksendal}. 
Here we shall only sketch a proof of the fact that the process
$\tilde{W}(t)$ is indeed distributed according to ${\cal
N}(0,\sqrt{t})$, as the standard Brownian motion. First recall from
(\ref{eq:pdfBMD}) that the probability density $f_P(\tilde{x},t)$ of
$\tilde{W}(t)$ under the original measure $P$ is
\begin{equation}
f_P(\tilde{x},t)=\frac{1}{\sqrt{2t}} \exp\left\{-\frac{(\tilde{x}-a t)^2}{2t}\right\}.
\label{eq:FP}
\end{equation}
Now according to (\ref{eq:fQ}) and (\ref{eq:Mt}) we have that
\begin{equation}
f_Q(\tilde{x},t)= e^{-a \tilde{x}+\frac{1}{2}a^2 t} f_P(\tilde{x},t). \label{eq:fQ1}
\end{equation}
Inserting (\ref{eq:FP}) into (\ref{eq:fQ1}) then yields
\begin{equation}
f_Q(\tilde{x},t)=\frac{1}{\sqrt{2t}} \; e^{-\tilde{x}^2/2t}, \label{eq:fQ2}
\end{equation}
which is precisely the pdf for the standard Brownian motion. Q.E.D.

\medskip

One of the main applications of change of measures is to eliminate the
drift in stochastic differential equations, so that with respect to
the new measure $Q$ the process is a martingale. The measure $Q$ is
then called an {\it equivalent martingale measure}.  Constructing the
equivalent martingale measure for an arbitrary SDE of the form
(\ref{eq:X}) is a rather complicated procedure
\cite{shiryaev}.  One important exception are linear SDE's where the
Girsanov theorem gives the measure transformation in an explicit form, as
shown below.

Consider the geometric Brownian motion discussed in
Sec.~\ref{sec:GMB}. For technical reasons \cite{oksendal}, let us
restrict ourselves to its finite-horizon version:
\begin{equation}
dS=\mu S dt + \sigma S dW, \quad t<T. \label{eq:GMB}
\end{equation}
where $\mu$ and $\sigma$ are positive constants. This equation can
then be rewritten as
\begin{equation}
dS=\sigma S \left( \frac{\mu}{\sigma} dt + dW\right)=\sigma S d\tilde{W}, \label{eq:dS3}
\end{equation}
where 
\begin{equation}
\tilde{W}_t=(\mu/\sigma) t + W_t, \quad t<T.\label{eq:WtilGMB}
\end{equation}
Now, according to Girsanov theorem, $\tilde{W}_t$ is a standard Brownian
motion with respect to the measure $Q$ given in (\ref{eq:fQ}) with
$a=\mu/\sigma$, and since the SDE (\ref{eq:dS3}) has no drift, its
solution $S_t$ is a martingale w.r.t.~the measure $Q$.

\subsection{The `efficiency symmetry'  and  the no-arbitrage 
condition as its `conservation law'}
\label{sec:symm}

The notion of an efficient market plays a crucial role in Finance.
Roughly speaking, in an efficient market all relevant information is
already reflected in the prices \cite{ingersoll}. This means, in
particular, that past prices give no additional information that is
not already contained in the current price.  In an efficient market
prices thus adjust immediately to the arrival of new
information. Since the content of future information and their effect
on prices are unknown, it should be impossible to make definite
predictions about future price based on the information available
today. Thus, the best prediction for the expected future price
(discounted to the present time) should be today's price. In other
words, in a efficient market `discounted prices' should be a
martingale. This intuitive definition of efficiency is formalized
below. [For technical reasons the results of this section will be
restricted to the discrete-time case.  Their extension to the
continuous-time framework, although possible to a large extent, is
more complicated and will not be attempted in the present notes; see,
e.g., \cite{shiryaev}.]

Consider a market formed by two assets $(B,S)$, where $B$ is our usual
{\it risk-free} free asset and $S$ is a {\it risky} asset. We suppose
that $S(t)$ follows a  stochastic process on a probability
space $(\Omega,{\cal F},P)$ endowed with a filtration $\{{\cal
F}_t\}_{t\ge0}$.

\begin{definition}
\label{def:effic}
Suppose the market $(B,S)$ operates at discrete time $t_n$,
$n=1,2,....$ This market is said to be efficient if there exists (at
least one) probability measure $Q$, equivalent to $P$, such that the
`discounted price' $\displaystyle\frac{S(t)}{B(t)}$ is a martingale
with respect to the measure $Q$, that is,
\begin{equation}
E^Q\left[\frac{S(t)}{B(t)}~\left| \, {\cal
F}_{t_0}\right. \right]=\frac{S(t_0)}{B(t_0)},
\quad {\rm for}\quad t_0\le t, \label{eq:effic}
\end{equation}
where $E^Q$ means expected value w.r.t.~the measure $Q$.
\end{definition}

The requirement of efficiency, as defined above, is somewhat
reminiscent of a symmetry principle in Physics. Indeed, we can recast
definition (\ref{eq:effic}) by saying that in a efficient
market there exists a `special measure Q' with respect to which
discounted prices are invariant under a sort of `time translation,'
in the following sense:
\begin{equation}
E^Q\left[\frac{S(t+T)}{B(t+T)}~\left |\, {\cal
F}_{t}\right. \right]=\frac{S(t)}{B(t)}, \label{eq:effic2}
\end{equation}
for any $T>0$. 

In Physics, symmetry principles are intimately connected with
conservation laws. (Recall, for instance, that the invariance of
Newton's law under time translation implies conservation of energy.)
It is thus only natural to ask whether the `efficiency symmetry' above
also leads to a `conservation law.' Perhaps not surprisingly, this is
indeed the case, as stated in the following theorem, which is
sometimes referred to as the {\it First Fundamental Theorem} of asset
pricing.


\begin{thm}
Suppose the market $(B,S)$ operates at discrete time $t_n$,
$n=1,2,....$ Then this market is efficient if and only if it is
arbitrage-free.
\end{thm}
\noindent (See \cite{shiryaev} for a proof.)

\medskip

Recall that absence of arbitrage means that any self-financing
portfolio with zero initial value, and with no chance of
becoming negative, will remain zero-valued for all subsequent
times. More precisely, the no-arbitrage condition says
that
\begin{equation}
V(0) = 0 \ \ \mbox{and}\ \ V(t)\ge0 \ \ a.s. \ \ \Longrightarrow \ \
V(t)=0 \ \ a.s.,
\end{equation}
where $a.s.$ means {\it almost surely}, i.e., with probability 1.  The
absence of arbitrage can thus be interpreted as a kind of
`conservation law' for the ``vacuum state'' of the market: if you start
in a state with zero initial money and do not take any
risks, then you remain at this state for all times.  I thus find it
quite interesting that the so-called ``no-free-lunch'' condition can
actually be seen as the conservation law associated with the
efficiency symmetry.

Another very important result is the {\it Second Fundamental Theorem}
of asset pricing, linking the completeness of a market to the
uniqueness of its equivalent martingale measure.

\begin{thm}
An arbitrage-free $(B,S)$-market is complete if and only if the
equivalent martingale measure $Q$ is unique.
\end{thm}
\noindent (See \cite{shiryaev} for a proof.)

\medskip

We already know that the Black-Scholes model is complete.  Below we
will calculate its equivalent martingale measure explicitly, and it
will become clear from the construction that it is indeed unique.

\subsection{Pricing derivatives with the equivalent martingale measure}

The notion of an equivalent martingale measure can be used to price
derivatives in a rather direct way, without having to solve a PDE.
The idea is that in an efficient economy all financial assets are 
martingales with respect to the equivalent martingale measure $Q$. More
precisely, if $F(S,t)$ is a contingent claim with maturity $T$ and
payoff function $\Phi(S(T))$ then from (\ref{eq:effic}) we have
\begin{equation}
\frac{F(S,t)}{B(t)}=E^Q_{t,S}\left[\frac{\Phi(S(T))}{B(T)}\right], \label{eq:EQF}
\end{equation}
where the subscripts $t,S$ denote that the expected value is taken at
present time $t$ and with current value $S$, i.e., conditional to the
information available at time $t$. It is not hard to convince oneself
that if the derivative price $F$ were not given by (\ref{eq:effic2}),
then there would be an arbitrage opportunity.

In the Black-Scholes model, the risk-free asset is a bank account with
fixed interest rate $r$, i.e., $B(t)=e^{rt}$, so that (\ref{eq:EQF})
becomes
\begin{equation}
F(S,t)=e^{-r(T-t)}E^Q_{t,S}\left[\Phi(S(T))\right],
\end{equation}
or
\begin{equation}
F(S,t)=e^{-r(T-t)}\int_0^\infty \Phi(S')f_Q(S',T;S,t)dS', \label{eq:EQF1}
\end{equation}
where $f_Q(S,t;S_0,t_0)$ denotes the probability density, under the
equivalent martingale measure $Q$, of the process $S(t)$ with initial
value $S(t_0)=S_0$. All that remains to be done now is to find the
equivalent martingale measure $Q$ for the Black-Scholes model. To do
this, consider the process
\begin{equation}
Z(t)=\frac{S(t)}{B(t)}=e^{-rt}S(t).
\end{equation}
We then have
\begin{eqnarray*}
dZ&=&-re^{-rt}Sdt+e^{-rt}dS\\&=&(\mu-r)Zdt+\sigma ZdW\\&=&\sigma Zd\tilde{W}, \label{eq:dZQ}
\end{eqnarray*}
where
\begin{equation}
\tilde{W}(t)=\left[(\mu-r)/\sigma\right]t+W(t). \label{eq:tilW}
\end{equation}
Now recall that in the Black-Scholes model the stock price follows a
geometric Brownian motion
\begin{equation}
dS=\mu S dt + \sigma S dW,
\end{equation}
which in terms of the process $\tilde{W}(t)$ given in (\ref{eq:tilW})
reads
\begin{equation}
dS=r S dt +\sigma S d\tilde{W}. \label{eq:dStil}
\end{equation}

From Girsanov theorem we know that there is a equivalent martingale
measure $Q$ that turns $\tilde{W}(t)$ into a Brownian motion. Equation
(\ref{eq:dStil}) then shows that w.r.t~the measure $Q$ the price
$S(t)$ follows a geometric Brownian motion with mean rate of return
equal to $r$.  The probability density $f_Q(S',T;S,t)$ can now be
obtained directly from (\ref{eq:logN}), by simply setting $\mu=r$ and
$S_0=S$. One then gets
\begin{equation}
f_Q(S',T;S,t)=\frac{1}{S'\sqrt{2\sigma^2 \tau}}
\exp\left\{-\frac{\left[\ln\left(\frac{S'}{S}\right)- (r- {1\over
2} \sigma^2)\tau\right]^2}{2\sigma^2\tau}\right\}, \label{eq:fQS}
\end{equation}  
where $\tau=T-t$.
Inserting (\ref{eq:fQS}) into (\ref{eq:EQF1}) we obtain
\begin{equation}
F(t,S)= {e^{-r(T-t)}\over\sqrt{2\pi\sigma^2(T-t)}}\int_{0}^\infty \Phi(S')e^{[\ln(S'/S)-(r-{1\over2}\sigma^2)(T-t)]^2} \frac{dS'}{S'}, \label{eq:FF}
\end{equation}
which is precisely the expression obtained for the generic solution of
the Black-Scholes equation given in (\ref{eq:fFTS}). In the case of a
European call option we have $\Phi(S')=\max(S'-K,0)$, which inserted
into (\ref{eq:FF}) yields, after some algebra, the Black-Scholes
formula (\ref{eq:fBS}).

It is interesting to notice that under the equivalent martingale
measure $Q$ the stock price in the Black-Scholes model follows a
geometric Brownian motion with the mean rate of return equal to the
risk-free interest rate $r$; see (\ref{eq:dStil}).  It is as if all
investors were {\it risk neutral}, in the sense they would be willing
to invest on a risky stock even though its expected return is just
what a risk-free bank account would yield. For this reason, the
pricing method based on the equivalent martingale measure is commonly
referred to as {\it risk neutral valuation}. Of course, actual
investors are {\it not} risk neutral. However, in an efficient market
there is a `special reference frame' where investors can be treated as
if they were indeed insensitive to risk.

\section{Beyond the Standard Model of Finance I: Non-Gaussian Distributions}

We have seen above that the Black-Scholes model is an elegant and
powerful theoretical construct: it is complete, efficient,
arbitrage-free, and Gaussian (in the sense that the stock returns are
normally distributed).  It is thus important to ask whether real
markets actually fit into this nice framework.

There are two main ways in which real markets may deviate from the
standard Black-Scholes model: i) the returns may not be normally
distributed or ii) there may exist long-memory effects on the time
series.  In this Section we will discuss the possibility that asset
prices may follow a non-Gaussian stable L\'evy process, while in the
next section we investigate whether financial data might exhibit
long-term memory.

\begin{figure}[t]
\begin{center}
\includegraphics*[width=.9\columnwidth]{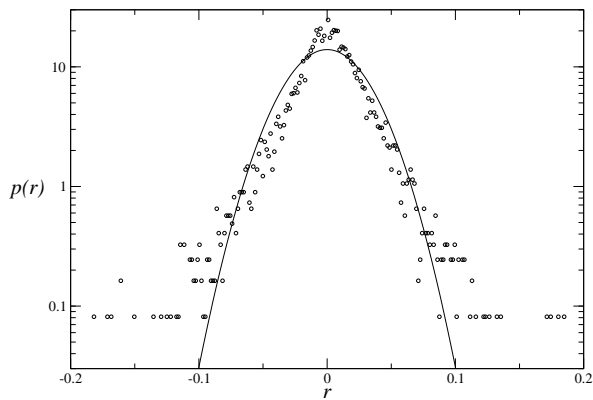}
\caption{Distribution of the daily Ibovespa returns (open circles). 
The solid line correspond to a Gaussian distribution with the 
same variance as that of the empirical distribution.}
\label{fig:hist}
\end{center}
\end{figure}

Mandelbrot \cite{mandelbrot} in 1963 was perhaps the first person to
challenge the paradigm that returns are normally distributed. He
analyzed cotton prices on various exchanges in the United States and
found evidences that their distribution of returns decays as a power
law and hence much slower than a Gaussian. An example of this `fat
tail' behavior can be seen in Fig.~\ref{fig:hist}, where I plot the
distribution for the returns of the Ibovespa index. In this figure, it
is also shown a Gaussian distribution with the variance of the
data, which appears as a parabola in the linear-log scale of the
graph. One clearly sees that the empirical distribution does indeed
have `fatter tails' when compared with the Gaussian distribution.

The Gaussian distribution is special for two main reasons.  The first
one is the Central Limit Theorem \cite{billingsley} that states that the
sum of infinitely many independent random variables (with finite
variance) will converge to a Gaussian variable. The second one is the
fact that it is a {\it stable} distribution, in the sense that the sum
of two independent Gaussian random variables is also a Gaussian
variable.  It is thus natural to ask whether there are other stable
distributions.  The French mathematician Paul L\'evy showed that there
is indeed a whole family of stable distributions of which the Gaussian
is but one particular case. In what follows, I will first
introduce the so-called L\'evy stable  distributions and then briefly
discuss their possible applications to financial data.

\subsection{The stable L\'evy distributions}

Let $X$ be a random variable with probability density function (pdf)
given by $p(x)$. We recall that the characteristic function $\varphi(z)$
of a random variable $X$ is the Fourier transform of its pdf $p(x)$:
\begin{equation}
\varphi(z)=\int_{-\infty}^\infty p(x) e^{izx} dx. \label{eq:varphi}
\end{equation}
Now let $X_1$ and $X_2$ be two independent random variables with pdf's
$p_1(x_1)$ and $p_2(x_2)$.  Since $X_1$ and $X_2$
are independent, the pdf of the sum $X=X_1+X_2$ is the convolution of the 
original pdf's:
\[
p(x)=\int_{-\infty}^\infty p_1(s)p_2(x-s)ds.
\]
Let now $\varphi(z,2)$ denote the characteristic function of $X$.
In view of the convolution theorem which states that the Fourier
transform of the convolution is the product of the Fourier transforms,
it then follows that
\begin{equation}
\varphi(z,2)=\varphi_1(z)\varphi_2(z). \label{eq:varz2}
\end{equation}
Suppose furthermore that $X_1$ and $X_2$ are  {\it identically
distributed}, that is,
\begin{equation}
X_1\stackrel{\rm d}=X_2,
\end{equation}
or alternatively
\begin{equation} 
\varphi_1(z)=\varphi_2(z)=\varphi(z).\label{eq:var1}
\end{equation}
(Recall that the symbol $\stackrel{\rm d}=$ denotes equality in the
distribution sense.)  From (\ref{eq:varz2}) and (\ref{eq:var1}) it
then follows that
\[
\varphi(z,2)=[\varphi(z)]^2.
\]

In general, if $X=\sum_{i=1}^N X_i$, where the $X_i$'s are independent
and identically distributed (i.i.d.) random variables, then
\begin{equation}
\varphi(z,N)=[\varphi(z)]^N \label{eq:varzN},
\end{equation}
from which the pdf of $X$ can be obtained by calculating the inverse
Fourier transform. Note that the pdf of the sum of $N$
i.i.d.~random variables will in general be quite different from the
pdf of the individual variables. There is however a special class of
distribution, the stable distributions, for which the pdf of the sum has
the same functional form of the individual pdf's.

\begin{definition}
A probability distribution $p(x)$ is stable if for each $N\ge2$ there
exist numbers $a_N>0$ and $b_N$, such that, if $X_1,...,X_N$ are
i.i.d.~random variables with distribution $p(x)$, then
\begin{equation}
X_1+...+X_N\stackrel{d}=a_nX_i+b_N. \label{eq:Xsum}
\end{equation}
\end{definition}

In other words, a distribution is stable if its form is invariant
under addition, up to a rescaling of the variable by a translation and
a dilation. More precisely, if $p(x,N)$ denotes the probability
density of $X=\sum_{i=1}^N X_i$, where the $X_i$'s are i.i.d.
variables with a stable distribution $p(x)$, then (\ref{eq:Xsum})
implies that
\begin{equation}
p(x,N)=\frac{1}{a_N} \, p\left(\frac{x -b_N}{a_N}\right). \label{eq:XN}
\end{equation}
Stability of probability distributions has a nice interpretation from
an economic standpoint. To see this, suppose that the variables $X_i$
represent daily increments of a given financial asset. Stability then
means {\it preservation} of the distribution under {\it time
aggregation}, which is a rather natural property to expect from
financial data.
 
As already mentioned, the Gaussian distribution is stable.  To see
this, recall that the Fourier transform of a Gaussian is a Gaussian
and that the product of Gaussians is again a Gaussian, so that from
(\ref{eq:varzN}) it follows that the characteristic function
$\varphi(z,N)$ will indeed be that of a Gaussian variable.  More
precisely, we have that the characteristic function of a normal
variable ${\cal N}(0, \sigma)$ is $\varphi(z)=e^{-(\sigma^2/2)z^2}$,
which inserted into (\ref{eq:varzN}) immediately yields
$\varphi(z,N)=e^{-(N
\sigma^2/2)z^2}$. Thus, the sum  of $N$ i.i.d.~normal
variables is normally distributed with standard deviation
$\sqrt{N}\sigma$, that is, $X\stackrel{d}= {\cal N}(0,
\sqrt{N}\sigma)$, which in turn implies that 
\begin{equation}
p(x,N)=\frac{1}{\sqrt{N}}\, p\left(\frac{x}{\sqrt{N}}\right).
\end{equation}
Thus, in the case of the Gaussian distribution we have $a_N=1/\sqrt{N}$
and $b_N=0$.

The class of stable distributions is rather small and was completely
determined by the mathematicians P.~L\'evy and
A.~Ya.~Khintchine in the 1920's. Here we shall restrict ourselves to
the subclass of symmetric distribution. In this case, the characteristic
function is given by
\begin{equation}
\varphi_\alpha(z)=e^{-a|z|^\alpha},   \label{eq:varlevy}
\end{equation}
where $0<\alpha\le2$ and $a>0$. The parameter $\alpha$ is called the
{\it stability exponent} and $a$ is a {\it scale} factor.  Taking
the inverse Fourier transform of $\varphi_\alpha(z)$ we 
obtain  the corresponding pdf $p_\alpha(x)$:
\begin{equation}
p_\alpha(x)=\frac{1}{2\pi}\int_{-\infty}^\infty \varphi_\alpha(z) e^{-izx} dz
=\frac{1}{\pi}\int_0^\infty e^{-az^\alpha} \cos(zx) dz. \label{eq:palpha}
\end{equation}
Unfortunately, however, only for two particular values of $\alpha$ can
the integral above be explicitly calculated:

\medskip

\noindent$\bullet$  $\alpha=1$ (Lorentzian or Cauchy distribution): 
$$p(x)=\frac{2a}{\pi}\frac{1}{x^2+4a^2}.$$

\medskip

\noindent$\bullet$ $\alpha=2$ (Gaussian distribution): 

$$p(x)=\frac{1}{8\pi a}e^{-x^2/4a}.$$

Note also that L\'evy distributions are not defined for $\alpha >2$,
because in this case the function obtained from (\ref{eq:palpha}) is
not everywhere positive.

Although, for arbitrary $\alpha$ the pdf $p_\alpha(x)$ cannot be found
in closed form, its asymptotic behavior for large $x$ can be easily
calculated from (\ref{eq:palpha}). Here one finds \cite{shiryaev} that
\begin{equation}
p_\alpha(x)\approx \frac{C_\alpha}{|x|^{1+\alpha}}, \quad |x|\to\infty, \label{eq:pasympt}
\end{equation}
where
\begin{equation}
C_\alpha=
\frac{a}{\pi} \; \Gamma(1+\alpha)\sin\frac{\pi\alpha}{2}.
\end{equation}
We thus see that the L\'evy distribution with $\alpha<2$ has the
interesting property that it shows {\it scaling} behavior for large
$x$, i.e., $p(x)$ decays as a power-law. 

Another important scaling relation for the L\'evy distribution can be
obtained, as follows. First note that for symmetric stable
distribution we necessarily have $b_N=0$. Now using (\ref{eq:varzN})
and (\ref{eq:varlevy}), we easily find that the dilation factor $a_N$
in (\ref{eq:XN}) is
\begin{equation}
a_N=N^{1/\alpha},
\end{equation}
so that (\ref{eq:XN}) becomes
\begin{equation}
p_\alpha(x,N)=\frac{p_\alpha\left(N^{1/\alpha}x\right)}{N^{1/\alpha}}, \label{eq:pxN}
\end{equation}
which implies  that
\begin{equation}
p(0,N)=\frac{p(0)}{N^{1/\alpha}}. \label{eq:P0N}
\end{equation}
One can then use this scaling relation to estimate the index
$\alpha$ of the L\'evy distribution: in a log-log plot of $p(0,N)$
against $N$, the slope of a linear fit gives precisely $1/\alpha$; see 
Sec.~\ref{sec:Levy} below.

The power-law decay of L\'evy distributions implies, of course, the
absence of a characteristic scale.  The downside of this is that all
L\'evy distributions have {\it infinite} variance! In fact, all
moments of order higher than 1 are infinite, since $E[|x|^n]$ diverges
for $n\ge\alpha$, as can be readily shown from
(\ref{eq:pasympt}). Processes with infinite variance are not
physically plausible, so several prescriptions to {\it truncate} the
L\'evy distribution at some large scale has been proposed, as
discussed next.

\subsection{Truncated L\'evy distributions}

To circumvent the problem of infinite variance in L\'evy
distributions, several truncation prescriptions have been proposed in
the literature. In a general  they can be written as
\begin{equation}
p(x)=p_\alpha(x)\Phi(x),
\end{equation}
where $\Phi(x)$ is a cut-off function to be chosen in such way that
the variance of the truncated distribution is finite. For example,
two possible choices that have been used to model the distributions of
financial asset prices are given below

\medskip
\noindent$\bullet$ Abruptly truncated L\'evy distribution (ATLD):

\[\Phi(x)=\Theta(x_c-|x|),\]
where $\Theta(x)$ is the Heaviside function and $x_c$ is some cut-off
length scale.

\medskip
\noindent$\bullet$ Exponentially truncated L\'evy distribution (ETLD):

\[\Phi(x)=A e^{-\lambda |x|},\]
where $\lambda>0$  and $A$ is a normalization factor.

Other variants of truncated L\'evy distributions  that have also been
considered in the literature are the gradually truncated L\'evy
distribution \cite{campanha} and the exponentially damped L\'evy
distributions \cite{gleria}.

Since a truncated L\'evy distribution has finite variance, then by
the central limit theorem the distribution of the sum $X=X_1+...+X_N$
of $N$ i.i.d~variables with such a distribution will converge to a
Gaussian distribution for large $N$. However, this convergence is
usually very slow---for financial data it is typically of the order of
tens of days; see below.  For shorter time scales, non-Gaussian behavior
may thus be of practical relevance.

\subsection{L\'evy distributions in Finance}
\label{sec:Levy}

L\'evy distribution have been used, for example, by Mantegna \&
Stanley \cite{MS} to model the distribution of changes in the stock
index S\&P500 of the American Stock Exchange. They analyzed
high-frequency data (one-minute quotes of the S\&P500) over the period
from January 1984 to December 1989. From the original time series
$Y(t)$ of the index values, they first generated time series
corresponding to index changes during intervals of $N$ minutes:
\begin{equation}
Z_N(t)\equiv Y(t+N)-Y(t).
\end{equation}
They then computed the empirical pdf $p(z,N)$ and analyzed the scaling
of $p(0,N)$ with $N$. In a log-log plot $p(0,N)$ showed a linear
behavior, as predicted by (\ref{eq:P0N}), with a slope corresponding
to $\alpha=1.4$ for $30<N<1000$ minutes \cite{MS}.  For $N>10^4$ the
slope of $p(0,N)$ approaches $-0.5$, indicating convergence to a
Gaussian behavior. In the L\'evy regime (i.e., small $N$), however,
the  tail of their empirical pdf decays slower than Gaussian but
faster than a pure L\'evy distribution with the exponent $\alpha$
found from above scaling argument.  These facts thus suggest that a
truncated L\'evy distribution would perhaps be more appropriate for
modeling the actual distribution.  Indeed, Bouchaud and Potters
\cite{BP} found that the probability of 15-minute changes of the
S\&P500 index is well described by a ETLD with $\alpha=1.5$.

The ETLD has also been applied by Miranda \& Riera \cite{rosane} to
study the daily returns of Ibovespa index of the S\~ao Paulo
Stock Exchange in the period 1986-2000. From the daily closing values
$Y(t)$ of the Ibovespa, they first calculated the time series for the
returns in intervals of $N$ days
\begin{equation}
r_N(t)=\log Y(t+N)-\log Y(t), \label{eq:rN}
\end{equation}
and then computed the corresponding pdf's for $p(r,N)$.  From the scaling
of $p(0,N)$ they found $\alpha\simeq 1.6-1.7$ for $N<20$ days, whereas
for larger $N$ a Gaussian-like behavior (i.e., $\alpha=0.5$) was
observed.

Many other applications of L\'evy processes in Finance have been
discussed in the literature \cite{bookLevy}. For example, a model for
option pricing has recently been considered where the price of the
underlying asset is assumed to follow a truncated L\'evy process
\cite{matacz}.  More recently, there have accumulated evidences
\cite{mccauley,yakovenko,mathia} that in certain cases financial data
may be better described by exponential distributions, rather than by
L\'evy or Gaussian distributions.

\section{Beyond the Standard Model of Finance II: Long-Range Correlations}

In this section, we discuss the possibility that asset prices might
exhibit long-range correlations and thus may need to be described in
terms of long-memory processes, such as the fractional Brownian
motion.

\subsection{Fractional Brownian motion}
\label{sec:FBM}

\begin{figure}[t]
\begin{center}
\includegraphics*[width=.9\columnwidth]{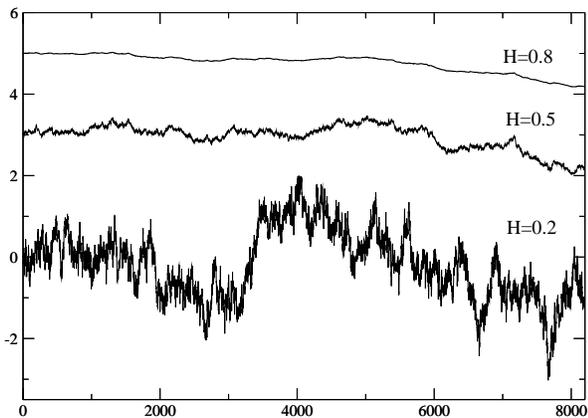}
\caption{Sample paths of fractional Brownian motion.}
\label{fig:FBM}
\end{center}
\end{figure}

The fractional Brownian motion (FBM) is a Gaussian process $\{W_H(t),
t>0\}$ with zero mean and stationary increments, whose variance and
covariance are given by
\begin{eqnarray}
&&E[W_H^2(t)]=t^{2H}, \label{eq:varBH}\\
&&E[W_H(s)W_H(t)]=\frac{1}{2}\left(s^{2H}+
t^{2H}-|t-s|^{2H}\right), \label{eq:corBH}
\end{eqnarray}
where $0<H<1$.  The FBM $W_H(t)$ is a self-similar process, in the
sense that
\begin{equation}
W_H(at)\stackrel{d}{=}a^HW_H(t),
\label{eq:aH}
\end{equation}
for all $a>0$.  A sample path of a FBM is therefore a fractal curve
with fractal dimension $D=1/H$. The parameter $H$ is called the
self-similarity exponent or the {\it Hurst exponent}. For $H=1/2$ the
process $W_H(t)$ corresponds to the usual Brownian motion, in which
case the increments $X_t=W_H(t+1)-W_H(t)$ are statistically
independent, corresponding to  white noise. On the other hand, for
$H\ne1/2$ the increments $X_t$, known as fractional white noise,
display long-range correlation in the sense that
\begin{equation}
E[X_{t+h}X_t]\simeq 2H(2H-1)h^{2H-2} \quad \mbox{for} \quad h\to\infty,
\end{equation}
as one can easily verify from (\ref{eq:varBH}) and (\ref{eq:corBH}).
Thus, if $1/2<H<1$ the increments of the FBM are positively correlated
and we say that the process $W_H(t)$ exhibits persistence. Likewise,
for $0<H<1/2$ the increments are negatively correlated and the FBM is
said to show antipersistence. Sample FBM paths with $H=0.2, 0.5,$ and
0.8 are shown in Fig.~\ref{fig:FBM}.

Several estimators for the exponent $H$ have been discussed in the
literature; see, e.g., Ref.~\cite{taqqu} for a comparison among some
of them.  One general methodology consists in estimating how the
`amount of fluctuation' within a time window of size $\tau$ scales
with $\tau$. Specific methods, such as the Hurst rescaled range (R/S)
analysis \cite{hurst} or the Detrendend Fluctuation Analysis
\cite{stanley1,jafferson}, differ basically on the choice of the
fluctuation measure. Here I shall discuss only the DFA that has proven
to be a more reliable estimator for $H$ than the Hurst R/S analysis
\cite{us}.

\subsection{Detrended fluctuation analysis}
 
Suppose we have a time series $r(t)$, $t=1,...,T$, corresponding to, say,
daily returns of a financial asset.  To implement the DFA,
we first integrate the original time series $r(t)$ to
obtain the cumulative time series $X(t)$:
\begin{equation}
X(t)=\sum_{t^\prime=1}^{t}(r(t^\prime)-\overline{r}), \quad t=1,...,T,
\end{equation}
where
\begin{equation}
\overline{r}=\frac{1}{T}\sum_{t'=1}^{T} r(t') .
\end{equation}
Next we break up $X(t)$ into $N$ non-overlapping time intervals,
$I_n$, of equal size $\tau$, where $n=0,1,...,N-1$ and $N$ corresponds
to the integer part of $T/\tau$. We then introduce the local trend
function $Y_\tau(t)$ defined by
\begin{equation}
Y_\tau (t)=a_n+b_nt \quad \mbox{for} \quad t\in I_n,
\end{equation}
where the coefficients $a_n$ and $b_n$ represent the least-square linear
fit of $X(t)$ in the interval $I_n$. Finally, we compute the
rescaled fluctuation function $F(\tau)$ defined as \cite{us}
\begin{equation}
F(\tau)=\frac{1}{S}\, \sqrt{\frac{1}{n\tau }\sum
_{t=1}^{N\tau}\left[X(t)-Y_\tau(t)\right]^2},
\label{eq:FS}
\end{equation}
where $S$ is the data standard deviation
\begin{equation}
S=\sqrt{\frac{1}{T}\sum _{t=1}^T\left(
r_{t}-\overline{r}\right)^{2}}.
\end{equation}
The Hurst exponent $H$ is then obtained from the scaling behavior of
$F(\tau)$:
\begin{equation}
F(\tau) = C \tau^H, 
\label{eq:alpha}
\end{equation}
where $C$ is a constant independent of the time lag $\tau$.

In a double-logarithmic plot the relationship (\ref{eq:alpha}) yields
a straight line whose slope is precisely the exponent $H$, and so a
linear regression of the empirical $F(\tau)$ will immediately give
$H$.  One practical problem with this method, however, is that the
values obtained for $H$ are somewhat dependent on the choice of the
interval within which to perform the linear fit
\cite{us,stanley2}.  It is possible to avoid part of this difficulty
by relying on the fact that for the fractional Brownian motion, the
fluctuation function $F(\tau)$ can be computed exactly \cite{taqqu}:
\begin{equation}
F_H(\tau) = C_H \tau^H, 
\label{eq:FH}
\end{equation}
where
\begin{equation}
C_H=\left[\frac{2}{2H+1}+\frac{1}{H+2}-\frac{2}{H+1}\right]^{1/2}.
\label{eq:CH}
\end{equation}
In (\ref{eq:FH}) we have added a subscript $H$ to the function
$F$ to denote explicitly that it refers to $W_H(t)$.  Equation
(\ref{eq:FH}) with (\ref{eq:CH}) now gives a one-parameter estimator
for the exponent $H$: one has simply to adjust $H$ so as to obtain the
best agreement between the theoretical curve predicted by $F_H(\tau)$
and the empirical data for $F(\tau)$.

\subsection{Fractional Brownian motion in Finance}

The idea of using the FMB for modeling asset price dynamics dates
back to the work of Mandelbrot \& van Ness \cite{MvN}.  Since then, the
Hurst exponent has been calculated (using different estimators) for
many financial time series, such as stock prices, stock indexes and
currency exchange rates \cite{peters,ausloos,pilar,tiziana,us}. In
many cases \cite{peters} an exponent $H>1/2$ has been found,
indicating the existence of long-range correlation (persistence) in
the data. It is to be noted, however, that the values of $H$ computed
using the traditional R/S-analysis, such as those quoted in
\cite{peters}, should be viewed with some caution, for this method
has been shown \cite{us} to overestimate the value of $H$. In this
sense, the DFA appears to give a more reliable estimates for $H$.

\begin{figure}[t]
\begin{center}
\includegraphics*[width=.9\columnwidth]{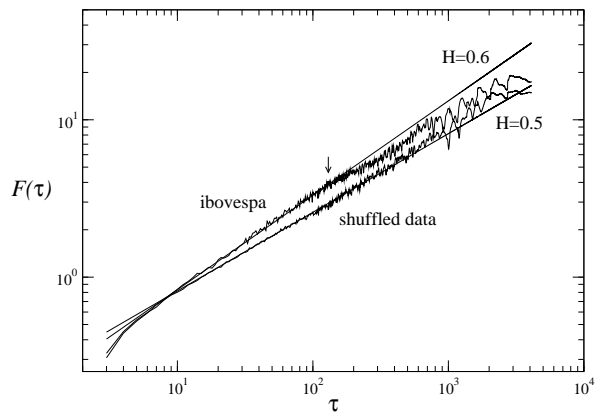}
\caption{Fluctuation function $F(\tau)$ as a function of  
$\tau$ for the returns of the Ibovespa index (upper curve) and for the
shuffled data (lower curve).  The upper (lower) straight  line gives
the theoretical curve $F_H(\tau)$ for $H=0.6$ ($H=1/2$).}
\label{fig:dfa}
\end{center}
\end{figure}

An example of the DFA applied to the returns of the Ibovespa stock
index is shown in Fig.~\ref{fig:dfa} (upper curve).  In this figure
the upper straight line corresponds to the theoretical curve
$F_H(\tau)$ given in (\ref{eq:FH}) with $H=0.6$, and one sees an
excellent agreement with the empirical data up to $\tau\simeq130$
days.  The fact that $H>0.5$ thus indicates persistence in the
Ibovespa returns. For $\tau>130$ the data deviate from the initial
scaling behavior and cross over to a regime with a slope closer to
$1/2$, meaning that the Ibovespa looses its `memory' after a period of
about 6 months.  Also shown in Fig.~\ref{fig:dfa} is the corresponding
$F(\tau)$ calculated for the shuffled Ibovespa returns. In this case
we obtain an almost perfect scaling with $H=1/2$, as expected, since
the shuffling procedure tends to destroys any previously existing
correlation.

As already mentioned, the Hurst exponent has been calculated for many
financial time series. In the case of stock indexes, the following
interesting picture appears to be emerging from recent studies
\cite{pilar,tiziana,us}: large and more developed markets, such as the
New York and the London Stock Exchanges, usually have $H$ equal to (or
slightly less than) 1/2, whereas less developed markets show a
tendency to have $H>1/2$.  In other words, large markets seem indeed
to be `efficient' in the sense that $H\simeq1/2$, whereas less
developed markets tend to exhibit long-range correlation. A possible
interpretation for this finding is that smaller markets are
conceivably more prone to `correlated fluctuations' and perhaps more
susceptible to being pushed around by aggressive investors, which may
explain in part a Hurst exponent greater than 1/2.

It should also be pointed out that a considerable time-variability of
the exponent $H$ for stock indexes has been found \cite{tiziana,us},
indicating that the data in such cases cannot be modeled in terms of
stationary stochastic processes. (A time-varying $H$ has also been
observed in other financial data, such as, currency exchange rate
\cite{ausloos}.) In such cases, the stationarity assumption is only a
rather crude approximation \cite{us}. Furthermore, the time dependence
of the Hurst exponent is an indication that the underlying process
might be multifractal rather than monofractal; see below.

\begin{figure}
\begin{center}
\includegraphics*[width=.9\columnwidth]{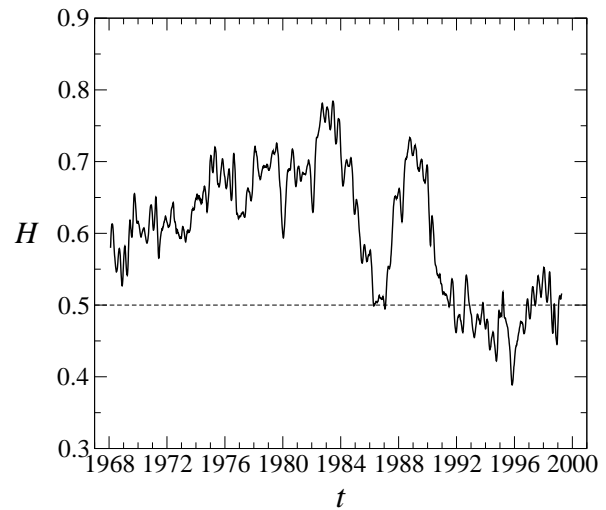}
\caption{The Hurst exponent $H$ for the Ibovespa as a function of time.
Here $H$ was computed in two-year time intervals, with the variable $t$
denoting the origin of each such interval.}
\label{fig:bianual}
\end{center}
\end{figure}

A time-varying Hurst exponent has been observed for the Brazilian stock
market. This case is of particular interest because in the recent past
Brazil was plagued by runaway inflation and endured several ill-fated
economic plans designed to control it.  To analyze the effect of
inflation and the economic plans, Costa and Vasconcelos \cite{us}
calculated a time-varying Hurst exponent for the Ibovespa returns,
computed in three-year time windows for the period 1968--2001. A
similar analysis but with two-year time windows is shown in
Fig.~\ref{fig:bianual}. One sees from this figure that during the
1970's and 1980's the curve $H(t)$ stays well above 1/2, the only
exception to this trend occurring around the year 1986 when $H$ dips
momentarily towards 1/2---an effect caused by the launch of the
Cruzado economic Plan in February 1986
\cite{us}.  In the early 1990's, after the launching of the Collor
Plan, we observe a dramatic decline in the curve $H(t)$ towards 1/2,
after which it remained (within some fluctuation) around $1/2$. This
fact has led Costa and Vasconcelos \cite{us} to conclude that the
opening and consequent modernization of the Brazilian economy that
begun with the Collor Plan resulted in a more efficient stock market,
in the sense that $H\simeq 0.5$ after 1990.  In Fig.~\ref{fig:bianual},
one clearly sees that after the launching of a new major economic
plan, such as the Cruzado Plan in 1986 and the Collor Plan in 1990,
the Hurst exponent decreases. This effect has, of course, a simple
economic interpretation: a Government intervention on the market is
usually designed to introduce ``anti-persistent effects,'' which in
turn leads to a momentary reduction of $H$.  Note also that only after
1990 does the curves $H(t)$ goes below 1/2.  This finding confirms the
scenario described above that more developed markets (as Brazil became
after 1990) tend to have $H\, {\stackrel{<}{\sim}} \, 1/2$.

\subsection{Option pricing under the FBM assumption}
\label{sec:OFBM}

We have seen above that often times asset prices have a Hurst exponent
different from 1/2. In such cases, the standard Black-Scholes model
does not apply, since it assumes that returns follow a Brownian motion
($H=0.5$).  A more appropriate model for the return dynamics would be
the fractional Brownian motion (FBM). Indeed, a `fractional
Black-Scholes model' has been formulated, in which it is assumed that
the stock price $S$ follows a geometric fractional Brownian motion
given by
\begin{equation}
dS=\mu S dt + \sigma S dW_H, \label{eq:dSH}
\end{equation}
where $W_H(t)$ is the standard FBM. The fractional stochastic
differential equation above is shorthand for the integral equation
\begin{equation}
S(t)=S(0) +\mu \int_0^t S(t') dt' + \sigma \int_0^tS(t') dW_H(t'). \label{eq:ISH}
\end{equation}
To make mathematical sense of this equation is, of course, necessary
to define stochastic integrals with respect to $W_H(t)$.  Here it
suffices to say that a fractional It\^o calculus can indeed be
rigorously defined \cite{HO} that shares (in an appropriate sense)
many of the properties of the usual It\^o calculus. In the context of
this fractional It\^o calculus is possible to prove that the solution
to (\ref{eq:dSH}) is given by
\begin{equation}
S(t)=S(0)\exp\left\{\mu t-\frac{1}{2}\sigma^2t^{2H}+\sigma W_H(t)\right\}. 
\end{equation}
Compare this expression with (\ref{eq:geoMBS}).

One can show \cite{HO} that the fractional Black-Scholes model is
complete and arbitrage-free.  To price derivatives with this model,
one can apply the same $\Delta$-hedging argument used before, which
now leads to the `fractional Black-Scholes equation.' The result is
summarized in the following theorem.

\begin{thm}
In the fractional Black-Scholes model, the price $F(S,t)$ of a
European contingent claim with payoff $\Phi(S(T))$ is given by the
solution to the following boundary-value problem
\begin{eqnarray}
{\partial F\over \partial t} + H\sigma^2 t^{2H-2}S^2{\partial^2
F\over \partial S^2} + rS{\partial F\over \partial S} - rF = 0,&&
\label{eq:FFBS}\\  F(T,S)=\Phi(S).
\end{eqnarray}
\end{thm}
[Compare with (\ref{eq:FBS}).]

For the case of a European call option the solution to the problem
above can be found in closed form. Alternatively, one can obtain the
option pricing formula directly from the equivalent martingale
measure, without having to solve the above PDE.  The final result for
this `fractional Black-Scholes formula' is given below \cite{necula}.

\begin{thm}
In the fractional Black-Scholes model, the price of a European call
option with strike price $K$ and maturity $T$ is given by
\begin{equation}
C(S,t) = S N(d_1) - K e^{-r(T-t)}N(d_2),
\label{eq:fC}
\end{equation}
where
\begin{eqnarray}
d_1 &=& \frac{\ln\left(S\over K\right) +
r(T-t)+ {1\over2}\sigma^2(T^{2H}-t^{2H})}{\sigma\sqrt{T^{2H}-t^{2H}}},\\ \\ d_2
&=& \frac{\ln\left(S\over K\right) +
r(T-t)-{1\over2}\sigma^2(T^{2H}-t^{2H})}{\sigma\sqrt{T^{2H}-t^{2H}}}.
\end{eqnarray} 
\end{thm}
[Compare with (\ref{eq:fBS}).]  

The fractional Black-Scholes formula has been applied to price some
options traded on the Brazilian market \cite{cajueiro}. Here, however,
the option prices obtained with the formula above resulted
considerably higher than those from the usual Black-Scholes
formula. The practical relevance of the fractional Black-Scholes model
to real markets thus needs to be investigated further.

\subsection{Multifractality in Finance}

The fact that the Hurst exponents of financial data often display
considerable variability in time indicates, as already mentioned, that
such time series cannot be satisfactorily modeled in terms of a
fractional Brownian motion, which is characterized by a constant $H$
and would thus capture only a sort of average behavior of the actual
price dynamics \cite{us}.  In such cases, it would be more appropriate
to model the data as a {\it multifractal} process.

The notion of a multifractal was first introduced in the context of
dynamical systems to describe physical processes taking place on a
fractal support \cite{ott}. A rigorous exposition of multifractal
measures is beyond the scope of the present notes, and so we will
content ourselves with a rather intuitive description of
multifractality. The basic idea here is that a {\it monofractal}
process, such as the FBM, is characterized by a single exponent $H$,
whereas for a multifractal a whole family (spectrum) of exponents is
necessary, one for each moment of the distribution.

We have seen above that the FBM is a Gaussian process whose standard
deviation scales with time as
\begin{equation}
\sqrt{E[W_H^2(t)]}=t^H.
\end{equation}
If we now introduce the generalized Hurst exponents $H_q$ as the
corresponding scaling exponent for the $2q$-th moment, that is,
\begin{equation}
\left\{E\left[W_H^{2q}(t)\right]\right\}^{1/2q}= C_qt^{H_q},
\end{equation}
where $C_q$ is a constant, it then immediately follows from property
(\ref{eq:Ex2n}) of the Gaussian distribution that
\begin{equation}
H_q=H. \label{eq:Hq}
\end{equation}
That is, all higher-order Hurst exponents of the FBM are equal to
$H$ itself and hence the FBM is said to be a monofractal.  Our working
definition of a {\it multifractal} will then be a process for which the
generalized Hurst exponents $H_q$ vary with $q$, or alternatively,
that the quantity $qH_q$ does not scale linearly with $q$.

In general, any method used to calculate the Hurst exponent $H$ (see
Sec.~\ref{sec:FBM}) can be adapted to obtain the generalized exponents
$H_q$. For example, the multifractal generalization of the DFA
consists in calculating the $q$th-order fluctuation function
$F_q(\tau)$,
\begin{equation}
F_q(\tau)= \left\{\frac{1}{N\tau}\sum _{t=1}^{N\tau}\left|X(t)-Y_\tau(t)\right|^{2q}
 \right\}^{1/2q}.
\label{eq:qFS}
\end{equation}
In complete analogy with (\ref{eq:alpha}), the exponents $H_q$ are
then obtained from the scaling
\begin{equation}
F_q(\tau)=C_q\tau^{H_q}. \label{eq:HqDFA}
\end{equation}
[We remark parenthetically that the multifractal DFA defined above is
slightly different from the formulation introduced in
Ref.~\cite{mdfa}, but such minor distinctions do not matter in
practice.]

\begin{figure}
\begin{center}
\includegraphics*[width=.9\columnwidth]{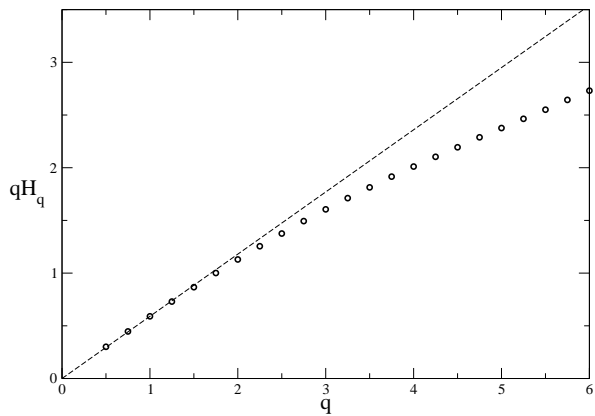}
\caption{The generalized Hurst exponents $H_q$ as a function of $q$
for the Ibovespa. The dashed line indicates the linear behavior 
of a monofractal FBM with $H=0.6$.}
\label{fig:multi}
\end{center}
\end{figure}

As an illustration of the multifractal DFA, we have applied this
method to the Ibovespa returns. For each $q$ we computed the function
$F_q(\tau)$ defined in (\ref{eq:HqDFA}), plotted it in a
double-logarithmic scale, and obtained the generalized exponent $H_q$
from the slope of the curve.  In Fig.~\ref{fig:multi} it is plotted
the resulting quantity $qH_q$ as a function of $q$. In this figure we
clearly see that $qH_q$ deviates from the linear behavior expected for
a monofractal, thus indicating that the time series of Ibovespa
returns does indeed display multifractal behavior. Evidences of
multifractal behavior have been seen in several other stock indexes
\cite{tiziana}.

Multifractality has also been observed in many time series for
currency exchange rates, and this has motivated the suggestion that
there might perhaps be a formal analogy between turbulent flows and
foreign currency exchange markets; see Ref.~\cite{voit} for references
to the original literature. In turbulence \cite{landau}, there is an
energy cascade from the large scales (where energy is fed into the
system) down to the small scales (where energy is dissipated by
viscosity).  In currency markets the analog would be a kind of {\it
information cascade} in time from long-term investors (that lake a
longer view of the market) to short-term investors (that watch the
market very frequently) \cite{voit}.

\section{Conclusions} 

In these notes, I tried to present a basic introduction to an
interdisciplinary area that has become known, at least among physicists
working on the field, as {\it Econophysics}.  I started out by giving
some basic notions about financial derivatives and illustrated how to
price them with a simple binomial model.  After this motivational
introduction, I offered a concise description of Brownian motion and
stochastic calculus, which provide the necessary mathematical tools to
describe financial asset prices in continuous time. I then proceeded to
discuss the Standard Model of Finance (SMF), namely, the Black-Scholes
model for pricing financial derivatives. The formulation of the
Efficient Market Hypothesis, which lies at the heart of the SMF, in
terms of martingales and its consequences for pricing derivatives were
also discussed. Finally, I briefly reviewed some recent work done
mostly, but not exclusively, by physicists that have produced evidences
that the SMF may not fully describe real markets. In this context,
some possible extensions of the Black-Scholes model were considered.

I should like to conclude by mentioning that other alternatives
approaches to the problem of pricing financial derivatives have been
proposed by physicists, using methods originally developed to treat
physical problems. For instance, the option pricing problem was recently
discussed in the context of the so-called non-extensive statistical
mechanics \cite{tsallis}. A ``Hamiltonian formulation'' for this
problem was also given in which the resulting ``generalized
Black-Scholes'' equation is formally solved in terms of path integrals
\cite{kleinert}. We refer the reader to Ref.~\cite{voit} for a brief
review of these and other  recent developments in Econophysics.

\acknowledgments I am grateful to Rog\'erio L.~Costa for many useful
discussions and for providing Fig.~\ref{fig:multi}. Financial support
from the Brazilian agencies CNPq and FINEP and from the special
research programs PRONEX is acknowledged.

\appendix
\section{Some Basic Definitions from Probability Theory}

\subsection{Probability space}

In probability theory one usually imagines performing an {\it
experiment} in which chance intervenes. The occurrence or nonoccurrence
of such an experiment is called an {\it outcome} $\omega$. The set of
all possible elementary outcomes is denoted by $\Omega$.  

An {\it event} $A$ is a set of outcomes, i.e., a subset of $\Omega$.
We are interested in attributing a probability to events in
$\Omega$. If $\Omega$ is finite or countable, we could introduce a
probability $P(\omega)$ for each individual outcome $\omega$ and then
define the probability $P(A)$ of an event $A$ as the sum of the
probabilities of all outcomes that make up the event $A$:
$$P(A)=\sum_{\omega\in A} P(\omega).$$ This procedure, however, will
not work when $\Omega$ is uncountable, i.e., $\Omega$ is a continuous
space such as ${\bf R}$, since in this case the probability of any
particular outcome is zero. Furthermore, a typical event will have
uncountably many (i.e. a continuum of) outcomes. Hence the formula
above is not applicable. That's why we need the notion of a
probability measure to be defined shortly. 

To do this, first we need to specify the class of `observable events',
i.e., the subsets of $\Omega$ to which a probability can be
associated.  If $\Omega$ is finite or countable, a partition of
$\Omega$ would be the natural candidate. (Recall that a {\it
partition} $\{A_i\}$ of a set $\Omega$ is a collection of disjoint
subsets, i.e., $A_i\subset \Omega$ and $A_i\cap A_j=\emptyset$ for $i\ne
j$, whose union covers the whole set $\Omega$, i.e., $\bigcup_i
A_i=\Omega$.) In the case of a continuous space this is not possible
and a different class of subsets is in order. To be useful, such a
class must be closed under the various set operations, such as union,
intersection, complementarity, etc. This is done through the concept
of a $\sigma$-algebra.

\begin{definition}
A family $\cal F$ of subsets of $\Omega$ is a $\sigma$-algebra on
$\Omega$ if the following conditions are fulfilled:
\begin{enumerate}
\item $\emptyset\in{\cal F}$ and $\Omega\in{\cal F}$
\item $A\in{\cal F} \Longrightarrow  A^c\in{\cal F}$, where $A^c=\Omega\setminus A$ is the complement 
of $A$ in $\Omega$
\item $A_1, A_2, ...\in {\cal F} \Longrightarrow \displaystyle\bigcup_{i=1}^\infty A_i \in {\cal F}$
\end{enumerate}
The par $(\Omega,{\cal F})$ is called a {\it measurable space}.
\end{definition}

The elements $A\subset {\cal F}$ of the $\sigma$-algebra $\cal F$ are
called {\it measurable sets} or simply {\it events}.  The idea here is
that for a given set $A\in{\cal F}$ it is possible to ascertain
whether any outcome $\omega$ belongs or not to $A$, and in this sense
the event $A$ is observable.

The smallest $\sigma$-algebra consists of the empty set $\emptyset$
and the set $\Omega$ itself, i.e., ${\cal F}_{\rm min}=\{\emptyset,
\Omega\}$.  The largest $\sigma$-algebra, on the other hand, is made
up of {\it all} subsets of $\Omega$, which is known as the power set
of $\Omega$ and denoted by $2^\Omega$, hence ${\cal F}_{\rm
max}=2^\Omega$.  Intermediate $\sigma$-algebras can be generated in
the following way. Start with a given family $\cal U$ of subsets of
$\Omega$ and form the intersection of all $\sigma$-algebras that
contain  $\cal U$:
\begin{equation}
{\cal F}_{\cal U}=\bigcap\{{\cal F}\; | \; {\cal F}\supset {\cal U}\}.
\end{equation}
In other words, ${\cal F}_{\cal U}$ is the smallest algebra that contains
$\cal U$ as is called the algebra generated by $\cal U$.

We can now attribute a `probability of occurrence' to events $A\in{\cal
F}$ via a probability measure on $(\Omega,{\cal F})$.

\begin{definition}
A probability measure $P$ on the measurable space $(\Omega,{\cal F})$
is a function $P:{\cal F}\to[0,1]$ such that
\begin{enumerate}
\item $P(\emptyset)=0$ and $P(\Omega)=1$
\item If $A_1, A_2, ...\in {\cal F}$ is a disjoint collection of elements 
of $\cal F$, i.e., $A_i U A_j=\emptyset$ if $i\ne j$, then $P(U_{i=1}^\infty A_i)=\sum_{i=1}^\infty P(A_i)$
\end{enumerate}
The triple $(\Omega,{\cal F},P)$ is called a {\it probability space}.
\end{definition}

\subsection{Random variables}

Intuitively, a random variable $X$ is a function that attributes to
each outcome $\omega$ a real number $x$, i.e., $X(\omega)=x$. We
usually think of $X$ as a random number whose value is determined by
the outcome $\omega$. A more formal definition is given in terms of
measurable functions.

\begin{definition}
Let $(\Omega,{\cal F},P)$ be a  probability space. A function
$f:\Omega\to {\bf R}$ is measurable with respect to the
$\sigma$-algebra $\cal F$, or more compactly, $\cal F$-measurable, if
$$f^{-1}(U)\equiv\{\omega\in\Omega | f(\omega)\in U\} \in {\cal F},$$
for all open sets $U\in{\bf R}$.
\end{definition}

The definition above means that for any given interval $(a,b)\subset {\bf R}$ there
is a meaningful event $A$ in $\Omega$. In an abuse of language we usually 
refer to this event as $A=\{a<f<b\}$.

\begin{definition}
A random variable $X$ on a probability space $(\Omega,{\cal F}, P)$ is
a  ${\cal F}$-measurable function.
\end{definition}

A random variable $X$ naturally generates a $\sigma$-algebra. This is
the algebra generated by all the sets $X^{-1}(U)$, $U\subset{\bf R}$
open, and is denoted ${\cal F}_X$. We think of ${\cal F}_X$ as
representing the `information' generated by the random variable $X$.
The $\sigma$-algebra ${\cal F}_X$ contains the essential information
about the structure of the random variable $X$; it contains all sets
of the form $\{\omega|a < X(\omega) < b\}$.

We also recall the definition of the probability distribution $F(x)$
of a random variable $X$:
\begin{equation}
F(x)=P(X\le x), \quad {\rm for}\quad x\in {\bf R}.
\end{equation}
Random variables can be either discrete, if the only assume a finite
or countably number of values $x_1, x_2, ...,$ or continuous.
Most continuous distributions of interest have a density $f(x)$,
i.e., a function $f(x)$ such that
\begin{equation}
F(x)=\int_{-\infty}^x f(x) dx.
\end{equation}
This allows us to compute the probability of a given event 
$A=\{a\le X\le b\}$ explicitly through the formula
\begin{equation}
P(A)= \int_a^b f(x) dx.
\end{equation}

\subsection{Stochastic processes}

Intuitively, a stochastic process represents a dynamical system which
evolve probabilistically in time. A formal definition is given below.

\begin{definition}
A stochastic process is a collection of random variables
$$\left\{X_t\right\}_{t\in T},$$ defined on some probability space
$(\Omega,{\cal F}, P)$ and parametrized by the variable $t$.
\end{definition}
 
We usually think of the label $t$ as being time, so that $X_t$ would
represent the (random) value of the quantity $X$, say, the price of a
risky asset or the position of a Brownian particle, at time $t$.  For
most of the cases, we consider $T$ to be the halfline $[0,\infty)$. In
this case we have a continuous-time process. Eventually, we shall also
consider discrete-time processes, in which case the variable $t$
assumes (non-negative) integer values, i.e., $T=\{0,1,2,...\}$.

It is perhaps worth emphasizing that a stochastic process is a
function of two variables. For a fixed time $t$, $X_t$ is a function
of the random variable $\omega$, i.e., $X_t=X_t(\omega)$.  For a fixed
outcome $\omega\in\Omega$, it is a function of time, $X_t=X_t(\omega),
\quad t\in T$. This function of time is called a {\it realization},
{\it path}, or {\it trajectory} of the stochastic process $X_t$.
Note, in particular, that in this context an {\it outcome} corresponds
an entire realization or trajectory of the stochastic process $X$.

A stochastic process is usually described in terms of the
distributions it induces.  The finite-dimensional distributions of the
stochastic process $X_t$ are the joint probability distributions
$p(x_1,t_1;...,x_n,t_n)$ of the random variables $X_{t_1},
...,X_{t_n}$, for all possible choices of times $t_1, ...,t_n\in T$
and every $n\ge 1$. The finite-dimensional distributions determine
many (but not all) relevant properties of a stochastic process. 
A special class of stochastic processes are the stationary ones.

\begin{definition}
A stochastic process is said to be {\it stationary} if all its
finite-dimensional distributions are invariant under a time translation,
that is,
$$p(x_1,t_1+\tau;...,x_n,t_n+\tau)=p(x_1,t_1;...,x_n,t_n),$$
for any $\tau>0$.  
\end{definition}
 
Another important class of stochastic process are Gaussian processes,
where all finite-dimensional distributions are (multivariate)
Gaussians.

\vspace{2cm}

\end{document}